\documentclass[11pt,a4paper]{article}
\usepackage[english]{babel}
\usepackage{amsmath,amsthm,amssymb,epsfig,latexsym}
\usepackage{color}
\usepackage{ulem}
\usepackage[numbers,square,sort&compress]{natbib}

\newcommand{\so}{\scriptscriptstyle \rm I}
\newcommand{\st}{\scriptscriptstyle \rm I\hspace{-1pt}I}

\newcommand{\bs}{\bar s}
\newcommand{\bt}{\bar t}

\newcommand{\bT}{\mathbb{T}}

\newcommand{\se}{\mathsf{e}}
\newcommand{\te}{\mathcal{J}}

\newcommand{\mb}[1]{\quad\mbox{ #1 }\quad}
\newcommand{\be}[1]{\begin{equation}\label{#1}}
\newcommand{\ba}[1]{\begin{multline}\label{#1}}
\newcommand{\ee}{\end{equation}}
\newcommand{\ea}{\end{eqnarray}}

\newcommand{\str}{\mathop{\rm str}}

\newtheorem{thm}{Theorem}[section]
\newtheorem{prop}{Proposition}[section]
\newtheorem{lemma}{Lemma}[section]

\def\qed{\hfill\nobreak\hbox{$\square$}\par\medbreak}
 \makeatletter
 \@addtoreset{equation}{section}
 \makeatother
 
\newcommand{\bea}{\begin{eqnarray}}
\newcommand{\eea}{\end{eqnarray}}

\newcommand{\bdr}{\mathbf{r}}

\begin{document}

\begin{flushright}
LAPTH-017/17
\end{flushright}

\vspace{12pt}

\begin{center}
\begin{LARGE}
{\bf Norm of Bethe vectors in models\\[2mm] with $\mathfrak{gl}(m|n)$ symmetry }
\end{LARGE}

\vspace{40pt}

\begin{large}
{A.~Hutsalyuk${}^{a,b}$,  A.~Liashyk${}^{c,d,e}$,
S.~Z.~Pakuliak${}^{a,f}$,\\ E.~Ragoucy${}^g$, N.~A.~Slavnov${}^h$\  \footnote{
hutsalyuk@gmail.com, a.liashyk@gmail.com, stanislav.pakuliak@jinr.ru, eric.ragoucy@lapth.cnrs.fr, nslavnov@mi.ras.ru}}
\end{large}

 \vspace{12mm}

${}^a$ {\it Moscow Institute of Physics and Technology,  Dolgoprudny, Moscow reg., Russia}

\vspace{4mm}

${}^b$ {\it Fachbereich C Physik, Bergische Universit\"at Wuppertal, 42097 Wuppertal, Germany}

\vspace{4mm}

${}^c$ {\it Bogoliubov Institute for Theoretical Physics, NAS of Ukraine,  Kiev, Ukraine}

\vspace{4mm}

${}^d$ {\it National Research University Higher School of Economics, Faculty of Mathematics, Moscow, Russia}

\vspace{4mm}

${}^e$ {\it Skolkovo Institute of Science and Technology, Moscow, Russia}

\vspace{4mm}

${}^f$ {\it Laboratory of Theoretical Physics, JINR,  Dubna, Moscow reg., Russia}

\vspace{4mm}

${}^g$ {\it Laboratoire de Physique Th\'eorique LAPTh, CNRS and USMB,\\
BP 110, 74941 Annecy-le-Vieux Cedex, France}

\vspace{4mm}

${}^h$  {\it Steklov Mathematical Institute of Russian Academy of Sciences, Moscow, Russia}

\end{center}

\vspace{4mm}


\begin{abstract}
We study quantum integrable models solvable by the nested algebraic Bethe ansatz and possessing
$\mathfrak{gl}(m|n)$-invariant $R$-matrix. We compute the norm of the Hamiltonian eigenstates.
Using the notion of a generalized model we show that the square of the norm obeys a number of
properties that uniquely fix it. We also show that a Jacobian of the system of Bethe equations
obeys the same properties. In this way we prove a generalized Gaudin hypothesis for the
norm of the Hamiltonian eigenstates.
\end{abstract}

\vspace{1cm}

\section{Introduction}

In 1972 M.~Gaudin formulated a hypothesis about the norm of the Hamiltonian eigenfunction of the quantum nonlinear Schr\"odinger equation
\cite{Gau72} (see also \cite{Gaud83}). According to this hypothesis, the square of the eigenfunction norm is proportional to a
Jacobian  closely related to the Bethe equations. In 1982 V. Korepin proved the Gaudin hypothesis for a wide class of quantum integrable
models \cite{Kor82}. In that work the Quantum Inverse Scattering Method (QISM) \cite{FadST79,FadT79,BogIK93L,FadLH96}  was used. An advantage
of this method is that it allows one to consider quantum models of different physical origin in a common framework. The work \cite{Kor82}
dealt with the models described by $\mathfrak{gl}(2)$-invariant $R$-matrix and its $q$-deformation. Using the same approach N. Reshetikhin
generalized this result to the models with $\mathfrak{gl}(3)$-invariant $R$-matrix \cite{Res86}. Recently, the norms of the Hamiltonian
eigenfunctions in the models with $\mathfrak{gl}(3)$ trigonometric $R$-matrix were calculated in \cite{Sla15}.

A new approach to the problem based on the quantized Knizh\-nik--Za\-mo\-lod\-chi\-kov equation was developed in a series of papers
\cite{VT,TarV96,MukV05}. There the norms of the eigenstates in $\mathfrak{gl}(N)$ based models were calculated. It was shown that these
results are equivalent to the Gaudin hypothesis. Concerning models described by superalgebras it is worth mentioning the work \cite{GomK99},
where an analog of the Gaudin formula was conjectured for Hubbard model.
Recently, the Gaudin norm of the full $\mathfrak{psu}(2,2|4)$ spin chain was studied
in \cite{BasCKLVZ}.

In all the cases listed above the original hypothesis was confirmed. Schematically it can be formulated as follows. Let $|\phi\rangle$ be
a Hamiltonian eigenstate. For quantum integrable models it can be parameterized by a set of parameters
$|\phi\rangle=|\phi(t_1,\dots,t_L)\rangle$ satisfying a system of equations (Bethe equations)
\be{BE-gen}
F_i(t_1,\dots,t_L)=1, \qquad i=1,\dots,L,
\ee
where $F_i$ are some functions depending on the model. Then the square of the norm of $|\phi\rangle$ is proportional to the following Jacobian
\be{Gaud-gen}
\langle\phi|\phi\rangle\sim\det \frac{ \partial \log F_i}{\partial t_j}.
\ee

In the present paper we prove the Gaudin hypothesis for integrable models with $\mathfrak{gl}(m|n)$ symmetry described by
the super-Yangian $Y\bigl(\mathfrak{gl}(m|n)\bigr)$. Our approach
is very closed to the one of the work \cite{Kor82}. It
is based on the  nested algebraic Bethe ansatz \cite{KulRes83,KulRes81,KulRes82} and the notion of a generalized model
\cite{Kor82,Kor82a,IzeK83} (see also \cite{BogIK93L}). We begin with  a {\it sum formula} for the scalar product of generic
Bethe vectors obtained in \cite{HutLPRS17b}. Using this formula we find a recursion for the scalar product and
then specify it to the case of the norm. In this way we prove that the norm and the Gaudin determinant satisfy the same recursion.
Taking into account the coincidence of the initial data, we thereby prove the Gaudin hypothesis for the models described
by  the super-Yangian $Y\bigl(\mathfrak{gl}(m|n)\bigr)$.

The paper is organized as follows. In section \ref{S-BN} we briefly recall basic notions of QISM specifying them
to the models based on  the super-Yangian $Y\bigl(\mathfrak{gl}(m|n)\bigr)$. In section~\ref{S-BV} we describe Bethe vectors of  the models with
$\mathfrak{gl}(m|n)$-invariant $R$-matrix
and consider their scalar products.
Section~\ref{S-GaudM} is devoted to the properties of the Gaudin matrix. Here we formulate the main result of the paper.
In section~\ref{S-GM} we introduce a notion of a generalized  model that serves a main tool of our approach. In section~\ref{S-QPSP} we find
a recursion for the scalar product of Bethe vectors. We specify this recursion to the case of the norm in section~\ref{S-Norm} and show that it coincides with the recursion for the Gaudin determinant. In this way we prove the generalized Gaudin hypothesis for the models with
$\mathfrak{gl}(m|n)$-invariant $R$-matrix. Several auxiliary statements are gathered in appendices.
In appendix~\ref{AS-RYY} we explain how to construct some representatives of the generalized model in the framework of evaluation representation.
Appendix~\ref{S-RHC} contains    recursions for the highest coefficients of the scalar products. Finally, in  appendix~\ref{SS-RP} we
find residues in the poles of the highest coefficients.

\section{Basic notions\label{S-BN}}

In this section we briefly recall basic notions of quantum integrable graded models. More detailed presentation
can be found in \cite{KulS80}.

The $\mathbb{Z}_2$-graded vector space $\mathbf{C}^{m|n}$ with the grading $[i]=0$ for $1\le i\le m$, $[i]=1$ for $m<i\le m+n$ is
a direct sum of spaces: $\mathbf{C}^{m|n}=\mathbf{C}^{m}\oplus \mathbf{C}^{n}$. Vectors belonging to $\mathbf{C}^{m}$ are called
even, vectors belonging to $\mathbf{C}^{n}$ are called odd. Matrices acting in $\mathbf{C}^{m|n}$ are  graded as $[E_{ij}]=[i]+[j]\in\mathbb{Z}_2$,
where $E_{ij}$ are elementary units: $(E_{ij})_{ab}=\delta_{ia}\delta_{jb}$.

The $R$-matrix of $\mathfrak{gl}(m|n)$-invariant models has the form
 \be{R-mat}
 R(u,v)=\mathbb{I}+g(u,v)P, \qquad g(u,v)=\frac{c}{u-v}.
 \ee
Here $c$ is a constant,  $\mathbb{I}$ and $P$ respectively are the identity matrix and the graded permutation operator \cite{KulS80}:
\be{EP}
\mathbb{I}=\mathbf{1}\otimes\mathbf{1}=\sum_ {i,j=1}^{n+m}E_{ii}\otimes E_{jj}, \qquad P=\sum_ {i,j=1}^{n+m}(-1)^{[j]}E_{ij}\otimes E_{ji}.
\ee
In \eqref{EP} we deal with the matrices acting in the tensor product $\mathbf{C}^{m|n}\otimes \mathbf{C}^{m|n}$.
In its turn, the tensor product of $\mathbf{C}^{m|n}$ spaces is  graded as follows:
\be{tens-prod}
(\mathbf{1}\otimes E_{ij})\,\cdot\,(E_{kl}\otimes \mathbf{1}) = (-1)^{([i]+[j])([k]+[l])}\,E_{kl}\otimes E_{ij}.
\ee

A basic relation of the QISM is an $RTT$-relation\footnote{Strictly speaking, in relation \eqref{RTT}, we should use
$R(u,v)\otimes\textsf{1}_{\mathcal{H}}$ instead of $R(u,v)$, where $\textsf{1}_{\mathcal{H}}$ is the unit acting on $\mathcal{H}$. This makes all relations very heavy, and we write loosely $R(u,v)$.
This will be the case throughout the paper, but we make this distinction in appendix \ref{AS-RYY} to clarify the construction of the evaluation map.}
\be{RTT}
R(u,v)\bigl(T(u)\otimes \mathbf{1}\bigr) \bigl(\mathbf{1}\otimes T(v)\bigr)= \bigl(\mathbf{1}\otimes T(v)\bigr)
\bigl( T(u)\otimes \mathbf{1}\bigr)R(u,v).
\ee
Here $T(u)$ is a monodromy matrix, whose matrix elements are quantum operators acting in a Hilbert space $\mathcal{H}$.
This Hilbert space coincides with the space of states  of the Hamiltonian  under consideration. The matrix elements
$T_{i,j}(u)$ are graded in the same way as the matrices $[E_{ij}]$: $[T_{i,j}(u)]=[i]+[j]\in\mathbb{Z}_2$.
Equation \eqref{RTT} holds in the tensor product $\mathbf{C}^{m|n}\otimes \mathbf{C}^{m|n}\otimes\mathcal{H}$.
All the tensor products are graded.

For given $R$-matrix \eqref{R-mat} the $RTT$-relation \eqref{RTT} implies a set of  commutation relations for the monodromy matrix entries
\begin{equation}\label{TM-1}
\begin{split}
[T_{i,j}(u),T_{k,l}(v)\}&
=(-1)^{[i]([k]+[l])+[k][l]}g(u,v)\Big(T_{k,j}(v)T_{i,l}(u)-T_{k,j}(u)T_{i,l}(v)\Big)\\
&=(-1)^{[l]([i]+[j])+[i][j]}g(u,v)\Big(T_{i,l}(u)T_{k,j}(v)-T_{i,l}(v)T_{k,j}(u)\Big),
\end{split}
\end{equation}
where we introduced the graded commutator
\be{Def-SupC}
[T_{i,j}(u),T_{k,l}(v)\}= T_{i,j}(u)T_{k,l}(v) -(-1)^{([i]+[j])([k]+[l])}   T_{k,l}(v)  T_{i,j}(u).
\ee
The Hamiltonian and other integrals of motion of a quantum integrable system can be obtained from a
graded transfer matrix. It is defined as the supertrace of the monodromy matrix
\be{transfer.mat}
\mathcal{T}(u)=\str T(u)= \sum_{j=1}^{m+n} (-1)^{[j]}\, T_{j,j}(u).
\ee
One can easily check \cite{KulS80} that $[\mathcal{T}(u)\,,\,\mathcal{T}(v)]=0$. Eigenstates of
the graded transfer matrix are eigenstates of the quantum Hamiltonian. As usual, they are defined
up to a normalization factor.  The main goal of this
paper is to find such normalization factors that the norms of the corresponding eigenstates are equal to $1$.

\section{Bethe vectors and their scalar products\label{S-BV}}

We do not specify  a Hilbert space $\mathcal{H}$ where the monodromy matrix entries act,
however, we assume that it contains a {\it pseudovacuum vector} $|0\rangle$, such that
 \be{Tij}
 \begin{aligned}
 &T_{i,i}(u)|0\rangle=\lambda_i(u)|0\rangle, &\qquad i&=1,\dots,m+n,\\
 & T_{i,j}(u)|0\rangle=0, &\qquad i&>j\,,
 \end{aligned}
 \ee
where $\lambda_i(u)$ are some scalar functions.  Below it will be convenient to deal with ratios of these functions
 \be{ratios}
 \alpha_i(u)=\frac{\lambda_i(u)}{\lambda_{i+1}(u)}, \qquad  i=1,\dots,m+n-1.
 \ee
In the framework of the generalized model  considered in this paper,
they remain free functional parameters. We discuss some properties of the generalized model
in section~\ref{S-GM}.

We also assume that the monodromy matrix entries act in a dual space $\mathcal{H}^*$ with a dual pseudovacuum $\langle0|$ such that
 \be{dTij}
 \begin{aligned}
 &\langle0|T_{i,i}(u)=\lambda_i(u)\langle0|, &\qquad i&=1,\dots,m+n,\\
 &\langle0| T_{i,j}(u)=0, &\qquad i&<j\,.
 \end{aligned}
 \ee
Here the functions $\lambda_i(u)$ are the same as in \eqref{Tij}.

In the framework of the algebraic Bethe ansatz, it is assumed that the  space of states $\mathcal{H}$ is generated by the action of
the upper triangular elements of the monodromy matrix $T_{i,j}(u)$ with $i<j$ onto the vector $|0\rangle$.
In physical models, vectors of the space $\mathcal{H}$  describe states with quasiparticles of different types (colors).
In  $\mathfrak{gl}(m|n)$-invariant models quasiparticles may have $N=m+n-1$ colors.
Let $\{r_1,\dots,r_N\}$ be a set of non-negative integers. We say that a state has coloring $\{r_1,\dots,r_N\}$,
if it contains $r_i$ quasiparticles of the color $i$, where $i=1,\dots,N$. The action of $T_{i,j}(u)$ onto a state of a fixed coloring creates $j-i$
quasiparticles of the colors $i,\dots,j-1$. More details on coloring can be found in \cite{HutLPRS17b}.

A Bethe vector is a polynomial in the creation operators $T_{i,j}$ with $i<j$ applied to the vector $|0\rangle$.
All the terms of this polynomial have the same coloring.
In this paper we do not use an explicit form of the Bethe vectors, however, the reader can find it
in \cite{HutLPRS17a}.  A generic Bethe vector of  $\mathfrak{gl}(m|n)$-invariant model depends on $N=m+n-1$ sets
of variables $\bt^1,\bt^2,\dots,\bt^N$ called Bethe parameters. We denote Bethe vectors by $\mathbb{B}(\bt)$, where
\begin{equation}\label{Bpar}
\bar{t}\
=\{t^{1}_{1},\dots,t^{1}_{r_{1}};
t^{2}_{1},\dots,t^{2}_{r_{2}};\dots;
t^{N}_{1},\dots,t^{N}_{r_{N}} \},
\end{equation}
and the  cardinalities $r_i$  of the sets $\bt^i= \{t^{i}_{1},\dots,t^{i}_{r_{i}}\}$ coincide with the coloring. Thus, each Bethe parameter $t^i_k$ can
be associated with a quasiparticle of the color $i$. We also introduce the total number of the Bethe parameters
\be{r}
{\bdr}=\#\bt=\sum_{i=1}^N r_i.
\ee

Bethe vectors are symmetric over permutations of the parameters $t^i_k$ within the set $\bt^i$, however, they are not symmetric over
permutations over parameters belonging to different sets $\bt^i$ and $\bt^j$. For generic Bethe vectors the
Bethe parameters $t^i_k$ are generic complex numbers. If these parameters satisfy a  special system of equations
(Bethe equations), then the corresponding vector becomes an eigenvector of the transfer matrix \eqref{transfer.mat}.
In this case it is called {\it on-shell Bethe vector}. We give explicitly the system of Bethe equations \eqref{BE}
a bit later, after introduction a necessary notation.

Dual Bethe vectors belong to the dual space $\mathcal{H}^*$. They can be obtained as a graded transposition of the Bethe vectors
(see e.g. \cite{PakRS17,HutLPRS17a,HutLPRS17b}). We denote dual Bethe vectors by $\mathbb{C}(\bt)$, where $\bt$ are the Bethe parameters
\eqref{Bpar}. Dual Bethe vectors become on-shell, if the set $\bt$ satisfy the system \eqref{BE}.

\subsection{Notation}

In this paper we use notation and conventions of the work \cite{HutLPRS17b}.
Besides the function $g(u,v)$  we use one more rational function
\be{fh}
f(u,v)=1+g(u,v)=\frac{u-v+c}{u-v}.
\ee
In order to make formulas uniform we also introduce a `graded' constant $c_{[i]}=(-1)^{[i]}c$. Respectively,
we use `graded' rational functions $g_{[i]}(u,v)$ and $f_{[i]}(u,v)$:
\be{desand}
\begin{aligned}
g_{[i]}(u,v)&=\frac{c_{[i]}}{u-v},\\
f_{[i]}(u,v)&=1+g_{[i]}(u,v)=\frac{u-v+c_{[i]}}{u-v}.
\end{aligned}
\ee
Finally, we define $\gamma_{i}(u,v)$ as
\be{gam-hg}
\gamma_{i}(u,v)=\left[\begin{array}{l}f_{[i]}(u,v),\qquad i\ne m,\\
g_{[i]}(u,v),\qquad i= m.
\end{array}\right.
\ee
Observe that the function $\gamma_i$ takes three values, namely, $\gamma_i(u,v)=f(u,v)$ for $i<m$, $\gamma_i(u,v)=g(u,v)$ for $i=m$,
and $\gamma_i(u,v)=f(v,u)$ for $i>m$.

Let us formulate now a convention on the notation. We use a bar to denote sets of variables.  The set of the Bethe parameters
is denoted by $\bt$ (like in \eqref{Bpar}) or $\bs$. The latter notation mostly is used for the Bethe parameters
of dual Bethe vectors. From now on individual Bethe parameters are labeled with a Greek superscript and a Latin subscript,
i.e. $t^\mu_j$, $t^\nu_k$, and so on. The superscript refers to the color, while the subscript counts the number of the
Bethe parameters of the fixed color. Thus, $\bt=\{\bt^1,\dots,\bt^N\}$, where
$\bt^\mu=\{t^\mu_1,\dots,t^\mu_{r_\mu}\}$. The integers $r_\mu$ denote the cardinalities $r_\mu=\#\bt^\mu$, and the
total cardinality ${\bdr}$ is given by \eqref{r}. Similar notation is used for the set $\bs$.

Below we consider partitions of the Bethe parameters into disjoint subsets. The subsets are denoted by Roman numbers, i.e.
$\bt^\mu_{\so}$, $\bs^\nu_{\st}$, and so on. A special notation $\bt^\mu_j$ (resp. $\bs^\mu_j$) is used for the
subset of $\bt^\mu$ (resp. $\bs^\mu$) complementary to the parameter $t^\mu_j$ (resp. $s^\mu_j$), i.e.
$\bt^\mu_j=\bt^\mu\setminus\{t^\mu_j\}$ (resp. $\bs^\mu_j=\bs^\mu\setminus\{s^\mu_j\}$).

We use a shorthand notation for products of  the  functions \eqref{ratios}, \eqref{desand}, and \eqref{gam-hg}.
 Namely, if some of these functions depends
on a set of variables (or two sets of variables), this means that one should take the product over the corresponding set (or
double product over two sets).
For example,
%
 %
 %
%
%
 \be{SH-prod}
 \alpha_\nu(\bt^\nu)= \prod_{t^\nu_j\in\bt^\nu} \alpha_\nu(t^\nu_j),\qquad\!\!
 f_{[\mu]}(t^{\mu}_k,\bt^\mu_k)=\prod_{\substack{t^\mu_\ell\in\bt^\mu\\ \ell\ne k}} f_{[\mu]}(t^{\mu}_k,t^\mu_\ell),\qquad\!\!
  \gamma_{\nu}(\bs^\nu_{\so},\bs^\nu_{\st})=\prod_{s^\nu_j\in\bs^\nu_{\so}}\prod_{s^\nu_k\in\bs^\nu_{\st}} \gamma_{\nu}(s^\nu_j,s^\nu_k).
 \ee

By definition, any product over the empty set is equal to $1$. A double product is equal to $1$ if at least one of the sets is empty.

To illustrate the use of the shorthand notation \eqref{SH-prod} we give here a system of
Bethe equations. Recall that if the Bethe parameters $\bt$ satisfy the system of Bethe equations, then the corresponding (dual) Bethe vector is on-shell.
Being written in a standard notation this system has the following form:
\be{BE00}
\alpha_\nu(t_j^\nu)=(-1)^{\delta_{\nu,m}(r_m-1)}\Biggl(\prod_{\substack{k=1\\k\ne j}}^{r_\nu}
\frac{\gamma_{\nu}(t_j^\nu,t_k^\nu)} {\gamma_{\nu}(t_k^\nu,t_j^\nu)}  \Biggr)
\frac{\prod_{k=1}^{r_{\nu+1}}f_{[\nu+1]}(t^{\nu+1}_k,t_j^\nu)}
{\prod_{k=1}^{r_{\nu-1}}f_{[\nu]}(t_j^\nu,t^{\nu-1}_k)}, \qquad
\begin{array}{l}\nu=1,\dots,N,\\ j=1,\dots,r_\nu.\end{array}
\ee
The use of the shorthand notation allows one to rewrite this system as
\be{BE}
\alpha_\nu(t_j^\nu)=(-1)^{\delta_{\nu,m}(r_m-1)}\frac{\gamma_{\nu}(t_j^\nu,\bt_j^\nu)f_{[\nu+1]}(\bt^{\nu+1},t_j^\nu)}
{\gamma_{\nu}(\bt_j^\nu,t_j^\nu)f_{[\nu]}(t_j^\nu,\bt^{\nu-1})}, \qquad \begin{array}{l}\nu=1,\dots,N,\\ j=1,\dots,r_\nu.\end{array}
\ee
The eigenvalue of the transfer matrix $\mathcal{T}(z)$ on the on-shell Bethe vector $\mathbb{B}(\bt)$ is then given by
\be{eigenval}
\tau(z|\bt)=\sum_{\nu=1}^{N+1}(-1)^{[\nu]}\lambda_\nu(z)f_{[\nu]}(z,\bt^{\nu-1}) f_{[\nu]}(\bt^{\nu},z).
\ee

\subsection{Initial normalization of Bethe vectors\label{SS-INBV}}

Although we do not use explicit formulas for the Bethe vectors, we should fix their initial normalization. We
use the same normalization as in \cite{HutLPRS17b}.

It was already mentioned  that
a generic Bethe vector has the form of a polynomial in $T_{i,j}$ with $i<j$ applied to the pseudovacuum $|0\rangle$. Among all the terms
of this polynomial there is one monomial that contains the operators $T_{i,j}$ with $j-i=1$ only. We call this monomial the {\it main term} and
fix the normalization of the Bethe vectors by fixing a numeric coefficient
of the main term
%
\be{mainterm-def}
\mathbb{B}(\bt)=\frac{ \bT_{1,2}(\bt^1)\dots \bT_{N,N+1}(\bt^N)|0\rangle}
{\prod_{i=1}^{N}\lambda_{i+1}(\bt^{i})\prod_{i=1}^{N-1}  f_{[i+1]}(\bt^{i+1},\bt^i)}
+\dots.
\ee
where ellipsis means all the terms containing at least one operator $T_{i,j}$ with $j-i>1$.
We also introduced symmetric operator products in \eqref{mainterm-def}:
\be{bT}
\bT_{i,i+1}(\bt^i)=\frac{T_{i,i+1}(t^i_1)\dots T_{i,i+1}(t^i_{r_i})}{\Bigl(\prod_{1\le j<k\le r_i}h(t^i_k,t^i_j)\Bigr)^{\delta_{i,m}}}.
\ee
One can easily check that due to the commutation relations \eqref{TM-1} the operator products $\bT_{i,i+1}(\bt^i)$ do are symmetric
over $\bt^i$ for all $i=1,\dots,m+n-1$.

Recall that we use here the shorthand notation for
the products of the functions $\lambda_{j+1}$ and $f_{[j+1]}$.
The  normalization in \eqref{mainterm-def} is different from the one used in \cite{HutLPRS17a} by the product $\prod_{j=1}^{N}\lambda_{j+1}(\bt^{j})$.
This additional normalization factor is convenient, because in this case the scalar products of the Bethe vectors depend
on the ratios $\alpha_i$ \eqref{ratios} only.

Since the operators $T_{i,i+1}$ and $T_{j,j+1}$ do not commute for $i\ne j$, the main term can be written in several forms corresponding to different
ordering of the monodromy matrix entries. The ordering  in \eqref{mainterm-def} naturally arises if we construct Bethe vectors via the embedding  of
$Y\bigl(\mathfrak{gl}(m-1|n)\bigr)$ into $Y\bigl(\mathfrak{gl}(m|n)\bigr)$.

\subsection{Scalar product of Bethe vectors\label{SS-SPBV}}

A scalar product of Bethe vectors is defined as
\begin{equation}\label{SP-def}
S(\bs|\bt)=\mathbb{C}(\bs) \mathbb{B}(\bt).
\end{equation}
Here  $\bs$ and $\bt$  are sets of generic complex numbers of the same cardinality $\#\bs=\#\bt$.
One can show that the scalar product of Bethe vectors of different coloring vanishes \cite{HutLPRS17b}, therefore, below
we consider only the case $\#\bs^\nu=\#\bt^\nu=r_\nu$, $\nu=1,\dots,N$ (recall  that $N=m+n-1$).

In \cite{HutLPRS17b} we found a sum formula for this scalar product
\begin{equation}\label{HypResh}
 S(\bs| \bt) = \sum
\frac{\prod_ {\nu=1}^{N}\alpha_{\nu}(\bs^\nu_{\so}) \alpha_{\nu}(\bt^\nu_{\st})  \gamma_{\nu}(\bs^\nu_{\st},\bs^\nu_{\so}) \gamma_{\nu}(\bt^\nu_{\so},\bt^\nu_{\st})}
{\prod_{\nu=1}^{N-1} f_{[\nu+1]}(\bar s^{\nu+1}_{\st},\bar s^\nu_{\so}) f_{[\nu+1]}(\bar t^{\nu+1}_{\so},\bar t^\nu_{\st})}
Z^{m|n}(\bs_{\so}|\bt_{\so}) \;\; Z^{m|n}(\bt_{\st}|\bs_{\st}).
\end{equation}
Here all the sets of the Bethe parameters $\bt^\nu$ and $\bs^\nu$ are divided into two subsets $\bt^\nu\Rightarrow\{\bt^\nu_{\so}, \bt^\nu_{\st} \}$
and $\bs^\nu\Rightarrow\{\bs^\nu_{\so}, \bs^\nu_{\st} \}$, such that $\#\bt^\nu_{\so}=\#\bs^\nu_{\so}$. The sum is taken over all possible partitions of this
type.

The function  $Z^{m|n}(\bs|\bt)$ is the highest coefficient (HC). This is a rational function of the Bethe parameters. It can be constructed recursively
starting with HC in $\mathfrak{gl}(1|1)$ superalgebra (see also \cite{HutLPRS16c} for an explicit determinant representation of HC in
$\mathfrak{gl}(2|1)$ superalgebra)
\be{HC11}
Z^{1|1}(\bs|\bt)=g(\bs,\bt).
\ee
The recursions for HC are given in appendix~\ref{S-RHC}.

The most important property of HC is that this function has  simple poles at $s_j^\mu=t_j^\mu$, $\mu=1,\dots,N$, $j=1,\dots,r_\mu$.

\begin{prop}\label{Res-HC}
 The residues of HC in the poles at $s_j^\mu=t_j^\mu$, $\mu=1,\dots,N$, $j=1,\dots,r_\mu$
are proportional to $Z^{m|n}(\bs\setminus \{s_j^\mu\}|\bt\setminus \{t_j^\mu\})$:
\begin{equation}\label{res-otv}
Z^{m|n}(\bs|\bt)\Bigr|_{s_j^\mu\to t_j^\mu}
= g_{[\mu+1]}(t_j^\mu,s_j^\mu)\,\frac{\gamma_{\mu}(\bt_j^\mu,t_j^\mu)\gamma_{\mu}(s_j^\mu,\bs_j^\mu) \;
Z^{m|n}(\bs\setminus \{s_j^\mu\}|\bt\setminus \{t_j^\mu\})}
{f_{[\mu+1]}(\bar t^{\mu+1},t^{\mu}_j)
f_{[\mu]}(s^{\mu}_j,\bar s^{\mu-1})}+reg,
\end{equation}
where $reg$ means regular terms.
\end{prop}

We prove this proposition in appendix~\ref{SS-RP}.

The square of the norm of the Bethe vector traditionally is defined as
\begin{equation}\label{2Norm-def}
S(\bt|\bt)=\mathbb{C}(\bt) \mathbb{B}(\bt),
\end{equation}
that is, this is the scalar product at $\bs=\bt$. Equation \eqref{HypResh} still holds in this case, however, separate terms of the sum
over partitions may have singularities due to the poles of HC. Thus, in order to approach the case of the norm one should take a limit $\bs\to\bt$ in \eqref{HypResh}. The limit $\bs\to\bt$ means that $s^\mu_j\to t^\mu_j$ for all $\mu=1,\dots,N$ and $j=1,\dots,r_\mu$.

Finally, to obtain the norm of on-shell Bethe vector, one should impose Bethe equations \eqref{BE}. According to the generalized
Gaudin hypothesis, the square of the norm of on-shell Bethe vector in $\mathfrak{gl}(m|n)$-invariant models is proportional to a special
Jacobian. We describe this Jacobian in the next section.

\section{Gaudin matrix\label{S-GaudM}}

The Gaudin matrix $G$ for  $\mathfrak{gl}(m|n)$-invariant models is an $N\times N$ block-matrix. The size of the block $G^{(\mu,\nu)}$ is $r_\mu\times r_{\nu}$.
To describe the entries $G^{(\mu,\nu)}_{jk}$ we introduce a function
\be{Phi}
\Phi^{(\mu)}_j= (-1)^{\delta_{\mu,m}(r_m-1)}\alpha_\mu(t^\mu_j)
\frac{\gamma_{\mu}(\bt_j^\mu,t_j^\mu)}{\gamma_{\mu}(t_j^\mu,\bt_j^\mu)}
\frac{f_{[\mu]}(t_j^\mu,\bt^{\mu-1})}{f_{[\mu+1]}(\bt^{\mu+1},t_j^\mu)}.
\ee
It is easy to see that Bethe equations \eqref{BE} can be written in terms of $\Phi^{(\mu)}_j$ as
\be{BE-log}
\Phi^{(\mu)}_j=1, \qquad \mu=1,\dots,N,\quad j=1,\dots,r_\mu.
\ee

The entries of the Gaudin matrix are defined as
\be{Glnjk}
G^{(\mu,\nu)}_{jk}=-c_{[\mu+1]}\frac{\partial\log\Phi^{(\mu)}_j}{\partial t^\nu_k}.
\ee

We are now in position to state the main result of this paper:
\begin{thm}
The square of the norm of the on-shell Bethe vector reads
\be{fin-answ0}
\mathbb{C}(\bt)\mathbb{B}(\bt)=\prod_{\nu=1}^{N}\prod_{\substack{p,q=1\\ p\ne q}}^{r_\nu}\gamma_{\nu}(t^\nu_p,t^\nu_q)
\left(\prod_{\nu=1}^{N-1} f_{[\nu+1]}(\bt^{\nu+1},\bt^{\nu})\right)^{-1} \det G,
\ee
where the matrix $G$ is given by \eqref{Glnjk}.
\end{thm}
We prove this formula in the rest of the paper.

\subsection{Properties of the Gaudin matrix\label{SS-PGM}}

First of all, let us give explicit expressions for the matrix elements of the Gaudin matrix \eqref{Glnjk}.
We have for the elements in the diagonal
blocks $G^{(\mu,\mu)}$:
\begin{multline}\label{Glljk}
G^{(\mu,\mu)}_{jk}=\delta_{jk}\Bigl[X_j^\mu-\sum_{\ell=1}^{r_{\mu}}\mathcal{K}_{\mu}(t_j^\mu,t_\ell^\mu)+
(-1)^{\delta_{\mu,m}}\sum_{q=1}^{r_{\mu-1}}\te_{[\mu]}(t_j^\mu,t_q^{\mu-1})\\
+\sum_{p=1}^{r_{\mu+1}}\te_{[\mu+1]}(t_p^{\mu+1},t_j^\mu)\Bigr]+\mathcal{K}_{\mu}(t_j^\mu,t_k^\mu).
\end{multline}
Here
\be{X}
X_j^\mu=-c_{[\mu+1]}\frac{d}{dz}\log\alpha_\mu(z)\Bigr|_{z=t_j^\mu},
\ee
and
\be{K}
\mathcal{K}_{\mu}(x,y)=\frac{2c^2(1-\delta_{\mu,m})}{(x-y)^2-c^2},\qquad\quad \te_{[\mu]}(x,y)=\frac{c^2}{(x-y)(x-y+c_{[\mu]})}.
\ee

The near-diagonal blocks are
\be{Gl1ljk}
G^{(\mu,\mu-1)}_{jk}=(-1)^{\delta_{\mu,m}+1}\te_{[\mu]}(t_j^\mu,t_k^{\mu-1}),
\qquad G^{(\mu,\mu+1)}_{jk}=-\te_{[\mu+1]}(t_k^{\mu+1},t_j^{\mu}).
\ee
If $|\mu-\nu|>1$, then $G^{(\mu,\nu)}_{jk}=0$.

Consider now some properties  of the Gaudin matrix determinant. Let
\be{F-detG}
\mathbf{F}^{({\bdr})}(\bar X; \bt)=\det G.
\ee
Here we have stressed that the function $\mathbf{F}^{({\bdr})}(\bar X; \bt)$ depends on two sets of variables. One of these  sets
consists of the Bethe parameters $\bt$ \eqref{Bpar}. Another set is
\begin{equation}\label{barX}
\bar{X}\
=\{X^{1}_{1},\dots,X^{1}_{r_{1}};
X^{2}_{1},\dots,X^{2}_{r_{2}};\dots;
X^{N}_{1},\dots,X^{N}_{r_{N}} \}.
\end{equation}
The superscript ${\bdr}$ shows the total number of  Bethe parameters or, what is the same, the total number of
parameters $X^\mu_j$: ${\bdr}=\#\bt=\#\bar X$.

In specific models the variables $X_j^\mu$ are functions of the Bethe parameters (see \eqref{X}). Here we consider a more general
case, where the sets $\bar X$ and $\bt$ are independent. In other words, we study $\det G$ with the matrix elements
\eqref{Glljk}, \eqref{Gl1ljk}, but we do not impose \eqref{X}.

\paragraph{Korepin criteria.}
The function $\mathbf{F}^{({\bdr})}(\bar X; \bt)$ obeys  some characteristic pro\-perties.
These properties listed below are quite analogous to the properties of the Gaudin determinant in the $\mathfrak{gl}(2)$ case.
Due to the parallel to the original paper \cite{Kor82} we call them {\it Korepin criteria}.


\begin{enumerate}
\item[(i)] The function $\mathbf{F}^{({\bdr})}(\bar X; \bt)$ is symmetric over the replacement of the pairs $(X_j^\mu,t_j^\mu)\leftrightarrow(X_k^\mu,t_k^\mu)$.
\item[(ii)] It is a linear function of each $X_j^\mu$.
\item[(iii)] $\mathbf{F}^{(1)}(X^1_1; t^1_1)=X^1_1$ for  $\#\bt={\bdr}=1$.
\item[(iv)] The  coefficient of  $X_j^\mu$ is given by a function
$\mathbf{F}^{({\bdr}-1)}$ with modified parameters $X_k^\nu$
\be{X-X}
\frac{\partial\mathbf{F}^{({\bdr})}(\bar X; \bt)}{\partial X_j^\mu}= \mathbf{F}^{({\bdr}-1)}(\{\bar{X}^{\text{mod}}\setminus X^{\text{mod};\mu}_j\};
\{\bt\setminus t^\mu_j\}),
\ee
where the original variables $X_k^\nu$ should be replaced by $X_k^{\text{mod};\nu}$:
\be{mad-X}
\begin{aligned}
&X_k^{\text{mod};\mu} =X_k^{\mu}-\mathcal{K}_{\mu}(t^\mu_j,t^\mu_k),\\
&X_k^{\text{mod};\mu+1} =X_k^{\mu+1}+(-1)^{\delta_{m,\mu+1}}\te_{[\mu+1]}(t^{\mu+1}_k,t^\mu_j),\\
&X_k^{\text{mod};\mu-1} =X_k^{\mu-1}+\te_{[\mu]}(t^\mu_j,t^{\mu-1}_k),\\
& X_{k}^{\text{mod};\nu}=X_{k}^\nu,\qquad |\nu- \mu|> 1.
\end{aligned}
\ee
\item[(v)] $\mathbf{F}^{({\bdr})}(\bar X; \bt)=0$, if all $X^\nu_j=0$.
\end{enumerate}

The properties (i)--(iv) are quite
obvious. In order to check the property (v) one should take the sum of all columns (or rows) of the matrix $G$
\be{sum-G}
\sum_{\nu=1}^{N}\sum_{k=1}^{r_\nu}G^{(\mu,\nu)}_{jk}=X^\mu_j.
\ee
Hence, if all $X^\mu_j=0$, then this linear combination vanishes, and thus, $\det G=0$.

\begin{prop}\label{Uni-KC}
The Korepin criteria fixes the function $\mathbf{F}^{({\bdr})}(\bar X; \bt)$ uniquely.
\end{prop}

{\sl Proof}.
The proof is exactly the same as in the $\mathfrak{gl}(2)$ case \cite{Kor82}. For completeness, we repeat it here.

Let functions $\mathbf{F}^{({\bdr})}_1(\bar X; \bt)$ and $\mathbf{F}^{({\bdr})}_2(\bar X; \bt)$ satisfy Korepin criteria.
Then for ${\bdr}=\#\bt=1$ we have $\mathbf{F}^{(1)}_1(X_1^1; t_1^1)=\mathbf{F}^{(1)}_2(X_1^1; t_1^1)$. Assume that
$\mathbf{F}^{({\bdr}-1)}_1(\bar X; \bt)=\mathbf{F}^{({\bdr}-1)}_2(\bar X; \bt)$. Then for $\#\bt={\bdr}$
we have
\be{difdif}
\frac{\partial}{\partial X^\mu_j}\bigl(\mathbf{F}^{({\bdr})}_1(\bar X; \bt)-\mathbf{F}^{({\bdr})}_2(\bar X; \bt))=0,
\ee
due to the property (iv) and the induction assumption, and
\be{dif0}
(\mathbf{F}^{({\bdr})}_1(\bar X; \bt)-\mathbf{F}^{({\bdr})}_2(\bar X; \bt))\Bigr|_{\bar X=0}=0,
\ee
due to the property (v). Since the function $\mathbf{F}^{({\bdr})}_1(\bar X; \bt)-\mathbf{F}^{({\bdr})}_2(\bar X; \bt)$ is linear over each $X^\mu_j$, equations \eqref{difdif} and
\eqref{dif0} yield $\mathbf{F}^{({\bdr})}_1(\bar X; \bt)-\mathbf{F}^{({\bdr})}_2(\bar X; \bt)=0$ for $\#\bt={\bdr}$.\qed

Thus, in order to prove \eqref{fin-answ0} it is enough to show that the properly normalized scalar product
of on-shell Bethe vectors $\mathbb{C}(\bt)\mathbb{B}(\bt)$ obeys Korepin criteria.


\section{Generalized model\label{S-GM}}

The notion of the generalized model was introduced in \cite{Kor82} for $\mathfrak{gl}(2)$ based models (see also
\cite{Kor82a,IzeK83,BogIK93L,Res86}). This model also can be considered in the case of the super-Yangian $Y\bigl(\mathfrak{gl}(m|n)\bigr)$.
In fact, the generalized model is a class of models. Each representative of this class has a monodromy matrix
satisfying the $RTT$-relation \eqref{RTT} with the $R$-matrix \eqref{R-mat}, and possesses
pseudovacuum vectors with the properties \eqref{Tij}, \eqref{dTij}. A representative of the generalized model can be characterized by
a set of the functional parameters $\alpha_\mu(u)$ \eqref{ratios}. Different representatives are distinguished by different
sets of the ratios $\alpha_\mu(u)$.

The sum formula \eqref{HypResh} for the scalar product is valid for any representative of the generalized model. Then we can consider the
scalar product as a function depending on two types of variables: the Bethe parameters $\bs$ and $\bt$ on the one hand, and the
functional parameters $\alpha_\mu$ on the other hand. Indeed, even if some $t^\mu_j$ (resp. $s^\mu_j$) is fixed, then the function $\alpha_\mu(t^\mu_j)$
(resp. $\alpha_\mu(s^\mu_j)$) changes freely
when running through the class of the generalized model. In particular,
using only inhomogeneous models with spins in higher dimensional representations one can easily construct
representatives of the generalized model (see appendix~\ref{AS-RYY}), for which
\be{alpha-rep}
\alpha_\mu(u)=\prod_{j=1}^{L^{(\mu)}} f_{[\mu]}(u,\xi^{(\mu)}_j).
\ee
Here inhomogeneities $\xi^{(\mu)}_j$ are arbitrary complex numbers, and $L^{(\mu)}$ are arbitrary positive integers. It is clear that even being restricted to this class of functions $\alpha_\mu$ we can approach any predefined value of $\alpha_\mu(u)$ at $u$ fixed.

The meaning of Bethe equations \eqref{BE} also changes in the generalized model. For a given representative
this is a set of equations for the Bethe parameters. In the generalized model this is a set of constraints between
two groups of independent variables $t^\mu_j$ and $\alpha_\mu(t^\mu_j)$. Indeed, one can fix an arbitrary set of the Bethe parameters $\bt$
and then find a set of functions $\alpha_\mu$ such that the system \eqref{BE} is fulfilled. For example, one can look for the functions $\alpha_\mu$
in the form \eqref{alpha-rep}. Then Bethe equations become a set of constraints for inhomogeneities $\xi^{(\mu)}_j$. Since the number
of inhomogeneities is not restricted, one can always provide solvability of the system \eqref{BE}.

We will see in section~\ref{S-QPSP} that
if  $t^\mu_j=s^\mu_j$ for some $\mu$ and $j$, then the scalar product depends also on the derivatives $\alpha'_\mu(t^\mu_j)$ of
the functional parameters $\alpha_\mu$.  They arise due to the presence of poles in the HC
$Z^{m|n}(\bs_{\so}|\bt_{\so})$ and $Z^{m|n}(\bt_{\st}|\bs_{\st})$.
The derivatives $\alpha'_\mu(t^\mu_j)$ also can be treated as independent functional parameters, because generically the values of a function
and its derivative in a fixed point are not related to each other. In particular, the square of the norm of a Bethe vector depends on
three type of variables: the Bethe parameters, the values of the functions $\alpha_\mu$ in the points $t^\mu_j$, and the values of the derivatives
$\alpha'_\mu$ in the same points. If the Bethe vector is on-shell, then we can express $\alpha_\mu(t^\mu_j)$ in terms of the Bethe parameters due to
\eqref{BE}. However, the derivatives $\alpha'_\mu(t^\mu_j)$ still remain free.  In particular, the variables $X^\mu_j$ \eqref{X} and the
Bethe parameters $\bt$ can be considered as independent variables in the framework of the generalized model.

To illustrate an advantage of the generalized model we prove here an identity that will be used below.

\begin{prop}\label{zero}
For arbitrary sets of complex parameters $\bt$ and $\bs$ (including $\bs = \bt$) such that  $\#\bs=\#\bt>0$,
we have the identity
\begin{equation}\label{SP-halp0}
\sum
\frac{\prod_ {\nu=1}^{N}  \gamma_{\nu}(\bs^\nu_{\so},\bs^\nu_{\st}) \gamma_{\nu}(\bt^\nu_{\st},\bt^\nu_{\so})}
{\prod_{\nu=1}^{N-1} f_{[\nu+1]}(\bar s^{\nu+1}_{\so},\bar s^\nu_{\st})
 f_{[\nu+1]}(\bar t^{\nu+1}_{\st},\bar t^\nu_{\so})}Z^{m|n}(\bs_{\so}|\bt_{\so}) \;\; Z^{m|n}(\bt_{\st}|\bs_{\st})=0.
\end{equation}
\end{prop}

{\sl Proof.} The lhs of equation \eqref{SP-halp0} is a rational function of the variables $\bs$ and $\bt$. Therefore, to show that it vanishes identically, it is enough
to prove that it is equal to zero for a finite number of sets $\bs$ and $\bt$. Consider one fixed pair of sets
$\bs$ and $\bt$  such that $\bs\ne\bt$. We choose a representative of the generalized model for which the
functions $\alpha_\nu(z)$ at the points $z=t_j^\nu$ are given by the rhs of equations \eqref{BE}, and at the points $z=s_j^\nu$ are given by the equations
\be{BEss}
\alpha_\nu(s_j^\nu)=(-1)^{\delta_{\nu,m}(r_m-1)}\frac{\gamma_{\nu}(s_j^\nu,\bs_j^\nu)f_{[\nu+1]}(\bs^{\nu+1},s_j^\nu)}
{\gamma_{\nu}(\bs_j^\nu,s_j^\nu)f_{[\nu]}(s_j^\nu,\bs^{\nu-1})}, \qquad \begin{array}{l}\nu=1,\dots,N,\\ j=1,\dots,r_\nu.\end{array}
\ee
Then $\mathbb{C}(\bs)$ and $\mathbb{B}(\bt)$ are different on-shell Bethe vectors for this representative of the generalized model. The corresponding eigenvalues
are $\tau(z|\bs)$ and $\tau(z|\bt)$  (see \eqref{eigenval}). Since $\bs\ne\bt$, these eigenvalues can coincide in a set of isolated points $z=z_i$, but not on the whole
complex plane. Thus, the scalar product $\mathbb{C}(\bs)\mathbb{B}(\bt)$ vanishes.

On the other hand, the scalar product of generic Bethe vectors is given by \eqref{HypResh}. In particular, we can apply this formula to the scalar product of the vectors considered above. Using the fact that $\bs\ne\bt$ we can express $\alpha_\nu(t_j^\nu)$ and $\alpha_\nu(s_j^\nu)$ respectively via \eqref{BE} and \eqref{BEss} and
substitute these expressions in \eqref{HypResh}. It is easy to see that then the scalar product $\mathbb{C}(\bs)\mathbb{B}(\bt)$ coincides with the lhs of equation \eqref{SP-halp0}. Thus, equation \eqref{SP-halp0} holds for a fixed pair $\bs$ and $\bt$. Finally, since the pair $(\bs,\bt)$ was arbitrary, by varying this pair,
we get an infinite number of points where the rational function
vanishes, hence the statement  of the proposition.. \qed

\textsl{Remark:} In the proof of Proposition \ref{zero}, we assumed that $\bs\ne\bt$ for convenience only, in order to avoid studying
the residues at the poles of the HC. We stress, however, that identity \eqref{SP-halp0} is valid for any complex parameters
$\bs$ and $\bt$, including $\bs=\bt$, because if a rational function is equal
to zero at a sufficiently large number of points, then it vanishes everywhere.

\section{Recursion for the scalar product\label{S-QPSP}}

Let us turn back to the scalar product in the form \eqref{HypResh}. Suppose that $s^\mu_j=t^\mu_j$ for some $j$ and $\mu$.
The total scalar product is not singular, because the $RTT$-commutation relations are not singular.
However, the highest coefficients in \eqref{HypResh} might have poles. The poles occur if either
$s^\mu_j\in \bs_{\so}$ and $t^\mu_j\in \bt_{\so}$ or  $s^\mu_j\in \bs_{\st}$ and $t^\mu_j\in \bt_{\st}$.
Resolving these singularities at $s^\mu_j=t^\mu_j$ we obtain derivatives of the
functions $\alpha_\mu(z)$. Our goal is to find, how the scalar product depends on these derivatives.

For this it  is convenient to introduce
\be{ha}
\begin{aligned}
\hat \alpha_\nu(t_j^\nu)&=(-1)^{\delta_{\nu,m}(r_m-1)}  \alpha_\nu(t_j^\nu)\frac{\gamma_{\nu}(\bt_j^\nu,t_j^\nu)f_{[\nu]}(t_j^\nu,\bt^{\nu-1})}
{\gamma_{\nu}(t_j^\nu,\bt_j^\nu)f_{[\nu+1]}(\bt^{\nu+1},t_j^\nu)},\\[4pt]
\hat \alpha_\nu(s_j^\nu)&= (-1)^{\delta_{\nu,m}(r_m-1)}   \alpha_\nu(s_j^\nu)\frac{\gamma_{\nu}(\bs_j^\nu,s_j^\nu)f_{[\nu]}(s_j^\nu,\bs^{\nu-1})}
{\gamma_{\nu}(s_j^\nu,\bs_j^\nu)f_{[\nu+1]}(\bs^{\nu+1},s_j^\nu)},
\end{aligned}
\qquad \begin{array}{l} \nu=1,\dots,N,\\ j=1,\dots, r_\nu,\end{array}
\ee
where (here and below) $\bt^0=\bs^0=\bt^{m+n}=\bs^{m+n}=\emptyset$. This implies in particular that the products involving elements
from  these empty sets  are equal to $1$.

Then, replacing $\alpha_\nu$ with $\hat\alpha_\nu$ in
the scalar product \eqref{HypResh} we arrive at
\begin{equation}\label{SP-halp}
 S(\bs| \bt) = \sum
\frac{ \prod_ {\nu=1}^{N}\hat\alpha_{\nu}(\bs^\nu_{\so}) \hat\alpha_{\nu}(\bt^\nu_{\st})
\gamma_{\nu}(\bs^\nu_{\so},\bs^\nu_{\st}) \gamma_{\nu}(\bt^\nu_{\st},\bt^\nu_{\so})}
{ \prod_ {\nu=1}^{N-1}f_{[\nu+1]}(\bs_{\so}^{\nu+1},\bs^{\nu}_{\st})f_{[\nu+1]}(\bt_{\st}^{\nu+1},\bt^{\nu}_{\so})}\;
 Z^{m|n}(\bs_{\so}|\bt_{\so}) \; Z^{m|n}(\bt_{\st}|\bs_{\st}).
\end{equation}
Note that the product of the sign factors $(-1)^{\delta_{\nu,m}(r_m-1)}$ gives $1$, because $\#\bs_{\so}^m +\#\bt_{\st}^m=r_m$.

Let $s^\mu_j\in \bs_{\so}$ and $t^\mu_j\in \bt_{\so}$. We denote the corresponding contribution to the scalar product by $S^{(1)}(\bs| \bt)$.
If $s^\mu_j\to t^\mu_j$, then due to \eqref{res-otv} the  HC $Z^{m|n}(\bs_{\so}|\bt_{\so})$ has a pole. Let
$\bs_{\so}^\mu=\{s^\mu_j,\bs^\mu_{\so'}\}$, $\bt^\mu_{\so}=\{t^\mu_j,\bt^\mu_{\so'}\}$, and  $\bs_{\so}^\nu=\bs^\nu_{\so'}$,
$\bt^\nu_{\so}=\bt^\nu_{\so'}$ for $\nu\ne\mu$. Then
using \eqref{res-otv} we obtain
\begin{equation}\label{res-HC1}
Z^{m|n}(\bs_{\so}|\bt_{\so})\Bigr|_{s_j^\mu\to t_j^\mu}
=g_{[\mu+1]}(t_j^\mu,s_j^\mu)\frac{\gamma_{\mu}(\bt^\mu_{\so'},t_j^\mu)\gamma_{\mu}(s_j^\mu,\bs_{\so'}^\mu) }
{f_{[\mu+1]}(\bar t^{\mu+1}_{\so},t^{\mu}_j)
f_{[\mu]}(s^{\mu}_j,\bar s^{\mu-1}_{\so})}\;Z^{m|n}(\bs_{\so'}\;|\;\bt_{\so'})+reg,
\end{equation}
where $reg$ means regular part.

The product of the $f$-functions and $\gamma$-functions in \eqref{SP-halp}  transforms as follows:
\begin{multline}\label{coeff1}
\frac{\prod_ {\nu=1}^{N}  \gamma_{\nu}(\bs^\nu_{\so},\bs^\nu_{\st}) \gamma_{\nu}(\bt^\nu_{\st},\bt^\nu_{\so})}
{\prod_{\nu=1}^{N-1} f_{[\nu+1]}(\bar s^{\nu+1}_{\so},\bar s^\nu_{\st})
 f_{[\nu+1]}(\bar t^{\nu+1}_{\st},\bar t^\nu_{\so})}=
 \frac{\gamma_{\mu}(s^\mu_j,\bs^\mu_{\st})   \gamma_{\mu}(\bt^\mu_{\st},t^\mu_j)}
{ f_{[\mu]}(s_j^\mu,\bs_{\st}^{\mu-1})  f_{[\mu+1]}(\bt_{\st}^{\mu+1},t_j^\mu)}\\
\times\frac{\prod_ {\nu=1}^{N}  \gamma_{\nu}(\bs^\nu_{\so'},\bs^\nu_{\st}) \gamma_{\nu}(\bt^\nu_{\st},\bt^\nu_{\so'})}
{\prod_{\nu=1}^{N-1} f_{[\nu+1]}(\bar s^{\nu+1}_{\so'},\bar s^\nu_{\st})
 f_{[\nu+1]}(\bar t^{\nu+1}_{\st},\bar t^\nu_{\so'})}.
\end{multline}
Combining \eqref{res-HC1} and \eqref{coeff1} we obtain for the contribution $S^{(1)}(\bs| \bt)$
\begin{multline}\label{SP-res1a}
S^{(1)}(\bs| \bt)\Bigr|_{s_j^\mu\to t_j^\mu} = \hat\alpha_\mu(s^\mu_j)g_{[\mu+1]}(t_j^\mu,s_j^\mu)
\frac{\gamma_{\mu}(\bt_{j}^\mu,t_j^\mu)  \gamma_{\mu}(s_j^\mu,\bs_{j}^\mu)}
{ f_{[\mu]}(s_j^\mu,\bs^{\mu-1})  f_{[\mu+1]}(\bt^{\mu+1},t_j^\mu)}\\
\times \sum
\frac{\prod_ {\nu=1}^{N} \hat\alpha_{\nu}(\bs^\nu_{\so'}) \hat\alpha_{\nu}(\bt^\nu_{\st})  \gamma_{\nu}(\bs^\nu_{\so'},\bs^\nu_{\st}) \gamma_{\nu}(\bt^\nu_{\st},\bt^\nu_{\so'})}
{\prod_{\nu=1}^{N-1} f_{[\nu+1]}(\bar s^{\nu+1}_{\so'},\bar s^\nu_{\st})
 f_{[\nu+1]}(\bar t^{\nu+1}_{\st},\bar t^\nu_{\so'})}
 Z^{m|n}(\bs_{\so'}|\bt_{\so'}) \; Z^{m|n}(\bt_{\st}|\bs_{\st})+reg,
\end{multline}
where now the sum is taken over partitions of the sets $\bt\setminus \{t_j^\mu\}$ and $\bs\setminus \{s_j^\mu\}$ respectively into subsets
$\{\bs_{\so'},\bs_{\st}\}$ and $\{\bt_{\so'},\bt_{\st}\}$. Recall also
that $\bs_{j}^\mu=\bs^\mu\setminus\{s_{j}^\mu\}$ and $\bt_{j}^\mu=\bt^\mu\setminus\{t_{j}^\mu\}$.

Similarly one can consider the case $s^\mu_j\in \bs_{\st}$ and $t^\mu_j\in \bt_{\st}$. Denoting the corresponding contribution
by $S^{(2)}(\bs| \bt)$ we find
\begin{multline}\label{SP-res2}
S^{(2)}(\bs| \bt)\Bigr|_{s_j^\mu\to t_j^\mu} = \hat\alpha_\mu(t^\mu_j)g_{[\mu+1]}(s_j^\mu,t_j^\mu)
\frac{ \gamma_{\mu}(\bs_{j}^\mu,s_j^\mu)\gamma_{\mu}(t_j^\mu,\bt_{j}^\mu)}
 {f_{[\mu]}(t_j^\mu,\bt^{\mu-1})f_{[\mu+1]}(\bs^{\mu+1},s_j^\mu)}\\
\times \sum
\frac{\prod_ {\nu=1}^{N} \hat\alpha_{\nu}(\bs^\nu_{\so}) \hat\alpha_{\nu}(\bt^\nu_{\st'})  \gamma_{\nu}(\bs^\nu_{\so},\bs^\nu_{\st'}) \gamma_{\nu}(\bt^\nu_{\st'},\bt^\nu_{\so})}
{\prod_{\nu=1}^{N-1} f_{[\nu+1]}(\bar s^{\nu+1}_{\so},\bar s^\nu_{\st'})
 f_{[\nu+1]}(\bar t^{\nu+1}_{\st'},\bar t^\nu_{\so})}
 Z^{m|n}(\bs_{\so}|\bt_{\so}) \; Z^{m|n}(\bt_{\st'}|\bs_{\st'})+reg.
\end{multline}
Here the sum is taken over partitions of the sets $\bt\setminus \{t_j^\mu\}$ and $\bs\setminus \{s_j^\mu\}$ respectively into subsets
$\{\bs_{\so},\bs_{\st'}\}$ and $\{\bt_{\so},\bt_{\st'}\}$.

Now we combine \eqref{SP-res1a} and \eqref{SP-res2}. Relabeling the subscripts of subsets $\so'\to\so$,  $\st'\to\st$
and substituting $\hat\alpha(s^\mu_j)$ and $\hat\alpha(t^\mu_j)$ respectively in terms of $\alpha(s^\mu_j)$
and $\alpha(t^\mu_j)$ we arrive at
\begin{multline}\label{SP-res}
S(\bs| \bt)\Bigr|_{s_j^\mu\to t_j^\mu} = g_{[\mu+1]}(t_j^\mu,s_j^\mu)\Bigl(\alpha_\mu(s^\mu_j)-\alpha_\mu(t^\mu_j)\Bigr)
 \frac{  (-1)^{\delta_{\mu,m}(r_m-1)}  \gamma_{\mu}(\bs_{j}^\mu,s_j^\mu) \gamma_{\mu}(\bt_j^\mu,t_{j}^\mu)}
 { f_{[\mu+1]}(\bs^{\mu+1},s_j^\mu)f_{[\mu+1]}(\bt^{\mu+1},t_j^\mu)}\\
\times \sum
\frac{\prod_ {\nu=1}^{N} \hat\alpha_{\nu}(\bs^\nu_{\so}) \hat\alpha_{\nu}(\bt^\nu_{\st})  \gamma_{\nu}(\bs^\nu_{\so},\bs^\nu_{\st}) \gamma_{\nu}(\bt^\nu_{\st},\bt^\nu_{\so})}
{\prod_{\nu=1}^{N-1} f_{[\nu+1]}(\bar s^{\nu+1}_{\so},\bar s^\nu_{\st})
 f_{[\nu+1]}(\bar t^{\nu+1}_{\st},\bar t^\nu_{\so})}
 Z^{m|n}(\bs_{\so}|\bt_{\so}) \; Z^{m|n}(\bt_{\st}|\bs_{\st})+\tilde S.
\end{multline}
Here $\tilde S$ denotes the terms that  depend on  the function $\alpha_\mu(t^\mu_j)$ but not on its derivative. The sum
is taken over partitions of the sets $\bt\setminus \{t_j^\mu\}$ and $\bs\setminus \{s_j^\mu\}$ respectively into subsets
$\{\bs_{\so},\bs_{\st}\}$ and $\{\bt_{\so},\bt_{\st}\}$.

Then performing the limit $ s_j^\mu\to t_j^\mu $ in \eqref{SP-res} we obtain
\begin{multline}\label{SP-ressolv}
S(\bs| \bt)\Bigr|_{s_j^\mu=t_j^\mu} =  (-1)^{\delta_{\mu,m}(r_m-1)}
\frac{X_j^\mu \alpha_\mu(t^\mu_j) \gamma_{\mu}(\bs_{j}^\mu,t_j^\mu) \gamma_{\mu}(\bt_j^\mu,t_{j}^\mu)}
 { f_{[\mu+1]}(\bs^{\mu+1},t_j^\mu)f_{[\mu+1]}(\bt^{\mu+1},t_j^\mu)}  \\
\times \sum
\frac{\prod_ {\nu=1}^{N} \hat\alpha_{\nu}(\bs^\nu_{\so}) \hat\alpha_{\nu}(\bt^\nu_{\st})  \gamma_{\nu}(\bs^\nu_{\so},\bs^\nu_{\st}) \gamma_{\nu}(\bt^\nu_{\st},\bt^\nu_{\so})}
{\prod_{\nu=1}^{N-1} f_{[\nu+1]}(\bar s^{\nu+1}_{\so},\bar s^\nu_{\st})
 f_{[\nu+1]}(\bar t^{\nu+1}_{\st},\bar t^\nu_{\so})}
 Z^{m|n}(\bs_{\so}|\bt_{\so}) \; Z^{m|n}(\bt_{\st}|\bs_{\st})+\tilde S,
\end{multline}
where $X_j^\mu$ is defined by \eqref{X} and $\tilde S$ does not depend on $X_j^\mu$.

One might have the impression that the sum over partitions in the second line of \eqref{SP-ressolv} gives the scalar product
$S(\bs\setminus \{s_j^\mu\}\;| \;\bt\setminus \{t_j^\mu\})$. This is not exactly so, because the functions $\hat\alpha_{\mu}$ and $\hat\alpha_{\mu\pm1}$ still
depend on  $t_j^\mu$ (see \eqref{ha}). However, we can get rid of this dependence if we introduce modified functional parameters $\alpha_{\nu}^{(\text{mod})}$. Namely, for $\mu$ fixed we set $\alpha_{\nu}^{(\text{mod})}(z)=\alpha_{\nu}(z)$, if $|\nu- \mu|> 1$, and
\be{mod-al}
\begin{aligned}
&\alpha_\mu^{(\text{mod})}(z) = (-1)^{\delta_{\mu,m}}   \alpha_\mu(z)\frac{\gamma_{\mu}(t^\mu_j,z)}{\gamma_{\mu}(z,t^\mu_j)},\\
&\alpha_{\mu+1}^{(\text{mod})}(z) =\alpha_{\mu+1}(z)f_{[\mu+1]}(z,t^\mu_j),\\
&\alpha_{\mu-1}^{(\text{mod})}(z) =\frac{\alpha_{\mu-1}(z)}{f_{[\mu]}(t^\mu_j,z)}.
\end{aligned}
\ee
Then, substituting $\hat\alpha_\nu$ in \eqref{SP-ressolv} in terms of $\alpha_{\nu}^{(\text{mod})}$ we obtain
\begin{multline}\label{SP-mod}
S(\bs| \bt)\Bigr|_{s_j^\mu=t_j^\mu} =  (-1)^{\delta_{\mu,m}(r_m-1)}
\frac{X_j^\mu\alpha_\mu(t^\mu_j) \gamma_{\mu}(\bs_{j}^\mu,t_j^\mu) \gamma_{\mu}(\bt_j^\mu,t_{j}^\mu)}
 { f_{[\mu+1]}(\bs^{\mu+1},t_j^\mu)f_{[\mu+1]}(\bt^{\mu+1},t_j^\mu)}  \\
\times \sum
\frac{\prod_ {\nu=1}^{N}\alpha_{\nu}^{(\text{mod})}(\bs^\nu_{\so}) \alpha_{\nu}^{(\text{mod})}(\bt^\nu_{\st})
\gamma_{\nu}(\bs^\nu_{\st},\bs^\nu_{\so}) \gamma_{\nu}(\bt^\nu_{\so},\bt^\nu_{\st})}
{\prod_{j=1}^{N-1} f_{[j+1]}(\bar s^{j+1}_{\st},\bar s^j_{\so})
 f_{[j+1]}(\bar t^{j+1}_{\so},\bar t^j_{\st})}Z^{m|n}(\bs_{\so}|\bt_{\so}) \;\; Z^{m|n}(\bt_{\st}|\bs_{\st})+\tilde S \, .
\end{multline}
The sum over partitions in \eqref{SP-mod} gives the scalar product $\mathbb{C}(\bs\setminus \{s_j^\mu\})
\mathbb{B}(\bt\setminus \{t_j^\mu\})$ in a new representative of the generalized model, in which the $\alpha$-functions are
modified according to \eqref{mod-al}.
Thus, we arrive at
\be{SP-SPmod}
S(\bs| \bt)\Bigr|_{s_j^\mu=t_j^\mu} = (-1)^{\delta_{\mu,m}(r_m-1)}
\frac{X_j^\mu\alpha_\mu(t^\mu_j) \gamma_{\mu}(\bs_{j}^\mu,t_j^\mu) \gamma_{\mu}(\bt_j^\mu,t_{j}^\mu)}
 { f_{[\mu+1]}(\bs^{\mu+1},t_j^\mu)f_{[\mu+1]}(\bt^{\mu+1},t_j^\mu)}
S^{(\text{mod})}(\bs\setminus \{s_j^\mu\}\;| \;\bt\setminus \{t_j^\mu\})+\tilde S,
\ee
where the modification of the scalar product means that now we should use the modified $\alpha$-functions \eqref{mod-al}.

Thus, we conclude that if $s_j^\mu=t_j^\mu$, then the scalar product linearly depends on the logarithmic derivative $X_j^{\mu}$. The
coefficient of $X_j^{\mu}$ is proportional to the modified scalar product $\mathbb{C}(\bs\setminus \{s_j^\mu\})
\mathbb{B}(\bt\setminus \{t_j^\mu\})$ in a new representative of the generalized model.

\section{Norm of on-shell Bethe vector\label{S-Norm}}

 It was already discussed that for $\bt=\bs$ the scalar product  depends on the
Bethe parameters $t_j^\nu$, the functional parameters $\alpha_\nu(t_j^\nu)$, and the  logarithmic
derivatives $X_j^\nu$ \eqref{X}.  In the case of the norm of on-shell Bethe
vectors the functions $\alpha_\nu$ are related to the parameters $\bt$ via Bethe equations \eqref{BE}.
Therefore, the norm of an on-shell Bethe vector
is a function of the Bethe parameters $t_j^\nu$ and the parameters $X_j^\nu$.

Let
\be{norm-funct}
\mathbf{N}^{({\bdr})}(\bar X; \bt) = \Biggl(\prod_{\nu=1}^{N}\prod_{\substack{p,q=1\\ p\ne q}}^{r_\nu}\gamma_{\nu}(t^\nu_p,t^\nu_q)\Biggr)^{-1}
\prod_{\nu=1}^{N-1} f_{[\nu+1]}(\bt^{\nu+1},\bt^{\nu}) \lim_{\bs\to\bt} \mathbb{C}(\bs)\mathbb{B}(\bt),
\ee
where $\mathbb{B}(\bt)$ is on-shell.

\begin{lemma}
The function $\mathbf{N}^{({\bdr})}(\bar X; \bt)$ fulfils the Korepin criteria.
\end{lemma}
{\sl Proof}.
Properties (i)--(ii) are quite obvious. Property (iii) follows from a direct calculation.
If only one Bethe parameter of the color $1$ is involved, then the Bethe
vector and the dual Bethe vector have respectively the following form  (see \cite{HutLPRS17a})
\be{BV11}
\mathbb{B}(t^1_1)=\frac{T_{1,2}(t^1_1)}{\lambda_2(t^1_1)}|0\rangle\,;\qquad
\mathbb{C}(t^1_1)=\langle0|\frac{T_{2,1}(t^1_1)}{\lambda_2(t^1_1)}.
\ee
Using commutation relations \eqref{TM-1} we immediately obtain
\be{SP11}
\mathbb{C}(s)\mathbb{B}(t)=\frac{\langle0|T_{2,1}(s)T_{1,2}(t)|0\rangle}{\lambda_2(s)\lambda_2(t)}=
(-1)^{[2]} g(s,t)\bigl(\alpha_{1}(t)-\alpha_{1}(s)\bigr).
\ee
Setting here $s=t=t^1_1$ we find
\be{Norm11}
\mathbb{C}(t^1_1)\mathbb{B}(t^1_1) =\alpha_{1}(t_1^1) X^1_1,
\ee
and finally, using the Bethe equation $\alpha_{1}(t_1^1)=1$ we arrive at property (iii).

The recursion \eqref{X-X} and the modification \eqref{mad-X}
follow from the considerations of the previous section. Indeed, taking the limit $\bs\to\bt$ in \eqref{SP-SPmod} we find
\be{LSP-SPmod}
\frac{\partial}{\partial X_j^\mu}\lim_{\bs\to\bt} S(\bs| \bt) = (-1)^{\delta_{\mu,m}(r_m-1)}  \alpha_\mu(t^\mu_j)
\left(\frac{ \gamma_{\mu}(\bt_{j}^\mu,t_j^\mu) }
 { f_{[\mu+1]}(\bt^{\mu+1},t_j^\mu)}\right)^2
\lim_{\bs\to\bt} S^{(\text{mod})}(\bs\setminus \{s_j^\mu\}\;| \;\bt\setminus \{t_j^\mu\}).
\ee
Substituting here $\alpha_\mu(t^\mu_j)$ from the Bethe equations \eqref{BE} we have
\be{L1SP-SPmod}
\frac{\partial}{\partial X_j^\mu}\lim_{\bs\to\bt} S(\bs| \bt) =
\frac{ \gamma_{\mu}(\bt_{j}^\mu,t_j^\mu) \gamma_{\mu}(t_j^\mu,\bt_{j}^\mu)}
 { f_{[\mu+1]}(\bt^{\mu+1},t_j^\mu)f_{[\mu]}(t_j^\mu,\bt^{\mu-1})}
\lim_{\bs\to\bt} S^{(\text{mod})}(\bs\setminus \{s_j^\mu\}\;| \;\bt\setminus \{t_j^\mu\}).
\ee
Thus, the coefficient of $\partial S/\partial X_j^\mu$ is proportional to the norm  of the Bethe vector of a new
representative of the generalized model. In this  representative the functional parameters
$\alpha_\nu$ should be modified according to \eqref{mod-al}. Obviously, this modification implies the modification \eqref{mad-X} of the
parameters $X_k^\mu$.

Remarkably, the new vector is still on-shell. Indeed, it is easy to see that the functional parameters $\alpha_\nu^{(\text{mod})}$
can be expressed in terms of the Bethe parameters $\bt\,\setminus\{t^\mu_j\}$ via Bethe equations. In particular,
\be{BE1}
\alpha_\mu^{(\text{mod})}(t_k^\mu)=(-1)^{\delta_{\mu,m}(r_m-2)}\frac{\gamma_{\mu}(t_k^\mu,\bt_{k,j}^\mu)f_{[\mu+1]}(\bt^{\mu+1},t_k^\mu)}
{\gamma_{\mu}(\bt_{k,j}^\mu,t_k^\mu)f_{[\mu]}(t_k^\mu,\bt^{\mu-1})},
\ee
where we introduced $\bt_{k,j}^\mu=\bt^\mu\setminus\{t_{j}^\mu,t_{k}^\mu\}$. Observe that if $\mu=m$, then $\#\bt_j^\mu=\#\bs_j^\mu=r_m-1$, therefore
the sign factor in \eqref{BE1} changes.
We also have
\be{BE2}
\begin{aligned}
\alpha_{\mu+1}^{(\text{mod})}(t_k^{\mu+1})&=(-1)^{\delta_{\mu+1,m}(r_m-1)}\frac{\gamma_{\mu+1}(t_k^{\mu+1},\bt_k^{\mu+1})f_{[\mu+2]}(\bt^{\mu+2},t_k^{\mu+1})}
{\gamma_{\mu+1}(\bt_k^{\mu+1},t_k^{\mu+1})f_{[\mu+1]}(t_k^{\mu+1},\bt^{\mu}_j)},\\
\alpha_{\mu-1}^{(\text{mod})}(t_k^{\mu-1})&=(-1)^{\delta_{\mu-1,m}(r_m-1)}\frac{\gamma_{\mu-1}(t_k^{\mu-1},\bt_k^{\mu-1})f_{[\mu]}(\bt^{\mu}_j,t_k^{\mu-1})}
{\gamma_{\mu-1}(\bt_k^{\mu-1},t_k^{\mu-1})f_{[\mu-1]}(t_k^{\mu-1},\bt^{\mu-2})}.
\end{aligned}
\ee
The other Bethe equations for $\alpha_{\nu}^{(\text{mod})}$ with $|\nu-\mu|>1$ do not change.
Thus, we arrive at the property (iv) for the function
$\mathbf{N}^{({\bdr})}(\bar X; \bt)$.

Finally, property (v) can be deduced as follows.  We take the general expression for the scalar product \eqref{HypResh}
and require that one of the sets, say $\bt$, satisfy the Bethe equations. Then we can replace the functions $\alpha_\nu(t^\nu_j)$ by the  expressions
in the rhs of the Bethe equations \eqref{BE}. After that, we consider the limit $\bs\to\bt$. Since all the poles of the HC in
\eqref{HypResh} are simple, it is enough to develop functions $\alpha_\nu(s^\nu_j)$ up to the first order over the difference
$s^\nu_j-t^\nu_j$:
\be{devel}
\alpha_\nu(s^\nu_j)=\alpha_\nu(t^\nu_j)+(s^\nu_j-t^\nu_j)\frac{d\alpha_\nu(z)}{dz}\Bigr|_{z=t^\nu_j}+O\left((s^\nu_j-t^\nu_j)^2\right).
\ee
If all $X^\nu_j=0$, then the derivatives of $\alpha_\nu$ vanish, and we can substitute $\alpha_\nu(s^\nu_j)$ given by \eqref{BEss} into \eqref{HypResh}
in the limit $\bs\to\bt$.
This leads us to
\begin{equation}\label{HypReshLim}
\lim_{\bs\to\bt} S(\bs| \bt) = \lim_{\bs\to\bt}\sum
\frac{\prod_ {\nu=1}^{N}\gamma_{\nu}(\bs^\nu_{\so},\bs^\nu_{\st}) \gamma_{\nu}(\bt^\nu_{\st},\bt^\nu_{\so})}
{\prod_{\nu=1}^{N-1} f_{[\nu+1]}(\bar s^{\nu+1}_{\so},\bar s^\nu_{\st})
 f_{[\nu+1]}(\bar t^{\nu+1}_{\st},\bar t^\nu_{\so})}Z^{m|n}(\bs_{\so}|\bt_{\so}) \;\; Z^{m|n}(\bt_{\st}|\bs_{\st}).
\end{equation}
However, due  to \eqref{SP-halp0} the sum over partitions in \eqref{HypReshLim} vanishes for arbitrary complex $\bs$ and $\bt$. In this way we arrive
at the property (v).\qed

Due to proposition~\ref{Uni-KC} we conclude that
\be{N-detG}
\mathbf{N}^{({\bdr})}(\bar X; \bt)=\det G,
\ee
leading to \eqref{fin-answ0}.

\section*{Conclusion}

We considered a generalized quantum integrable model with $\mathfrak{gl}(m|n)$-invariant $R$-matrix. We showed that the
square of the norm of on-shell Bethe vectors of this model is proportional to a Jacobian of the system of Bethe equations.
This result completely matches to the original Gaudin hypothesis on the norm of the Hamiltonian eigenvector. One can expect
that this hypothesis can be further generalized. In particular, it is quite natural to have a similar formula for the
models based on $U_q(\widehat{\mathfrak{gl}}(m))$ and $U_q(\widehat{\mathfrak{gl}}(m|n))$ algebras.
This will be the subject of our further publications.

The problem of the norm of on-shell Bethe vectors is very important for the calculation of form factors and correlation
functions in the models of physical interest. Further development in this direction requires more detailed analysis of
the Bethe vectors scalar products. Formally, the sum formula   gives an explicit result for the scalar product of generic Bethe
vectors, however, this representation is not convenient for applications in many cases. At the same time, one can hope
to find more compact representations for particular cases of the scalar product, as it was done in the models
with $\mathfrak{gl}(2|1)$-symmetry \cite{HutLPRS16b}. At present, work in this direction is underway.

\section*{Acknowledgements}
The work of A.L. has been funded by Russian Academic Excellence Project 5-100, by Young Russian Mathematics
award and by joint NASU-CNRS project F14-2017. The
work of S.P. was supported in part by the RFBR grant 16-01-00562-a.

\appendix

\section{$Y\bigl(\mathfrak{gl}(m|n)\bigr)$ representations induced  from $\mathfrak{gl}(m|n)$ ones\label{AS-RYY}}
A wide class of representations for the Yangian $Y\bigl(\mathfrak{gl}(m|n)\bigr)$  can be constructed from representations of $\mathfrak{gl}(m|n)$.
The construction relies on the notion of evaluation morphism and evaluation representations  \cite{CharyPress,Nazarov}.
Before detailing it, we make a short summary on irreducible representations of $\mathfrak{gl}(m|n)$.
\subsection{Highest weight representations of the Lie superalgebra $\mathfrak{gl}(m|n)$}
For simplicity, we present highest weight representations for the Lie superalgebra $\mathfrak{gl}(m|n)$ with $m\neq n$, but most of the discussion applies also to the case $m=n$.
Highest weight representations were studied in \cite{Kac78,HurniMorel}, see also \cite{dico} for a review on superalgebras.
We introduce the  $ \mathfrak{gl}(m|n)$  generators $\se_{ij}$ obeying
\begin{equation}\label{com-glmn}
[ \se_{ij}\,,\,\se_{kl}\} = \delta_{kj}\,\se_{il} - (-1)^{([i]+[j])([k]+[l])}\,\delta_{il}\,\se_{kj}.
\end{equation}

Highest weight representations of the Lie superalgebra  $\mathfrak{gl}(m|n)$ are characterized by a weight
$\boldsymbol{\lambda}=(\lambda_1,...,\lambda_{m+n})\in\mathbf{C}^{m+n}$ and a highest weight vector $|0\rangle$ such that
\be{hwglmn}
e_{ii}|0\rangle=\lambda_i|0\rangle \mb{and} e_{ij}|0\rangle=0,\quad i<j,
\ee
where $e_{ij}$ are the representatives of the $\mathfrak{gl}(m|n)$ generators.
 The highest weight vector $|0\rangle$ will produce the pseudovacuum \eqref{Tij}
through the evaluation morphism, see section \ref{sec:eval} below.
In other words, if $\pi_{\boldsymbol{\lambda}}$ denotes the mapping from the superalgebra to a representation space $\cal V_{\boldsymbol{\lambda}}$, then
$e_{ij}=\pi_{\boldsymbol{\lambda}}(\se_{ij})$ is a matrix (or an operator for infinite dimensional representations) acting on vectors in $\cal V_{\boldsymbol{\lambda}}$.
The associated Kac module is  obtained through the (multiple) applications of the representatives $e_{ij}$, $i>j$, on $|0\rangle$.

Among highest weight representations, the finite dimensional ones are characterized\footnote{For superalgebras, the irreducible part of the representation can be a coset of the Kac module, due to the existence of atypical representations.} by \textit{integrable dominant weights}, such that
$$
\lambda_{i}-\lambda_{i+1}\in\mathbf{Z}_+\,,\ i\neq m,\quad1\leq i\leq m+n-1
\mb{and} \lambda_m\in\mathbf{R}.
$$

Obviously any weight $\boldsymbol{\lambda}$ is a linear combination of the fundamental (dominant) weights\footnote{The last weight $\boldsymbol{\lambda}^{(m+n)}$ provides a trivial representation for $\mathfrak{sl}(m|n)$ and is related to the $\mathfrak{gl}(1)$ algebra which is central in $\mathfrak{gl}(m|n)$.}
$$
\boldsymbol{\lambda}^{(i)}=(\underbrace{1,...,1}_{i},\underbrace{0,...,0}_{m+n-i})\,,\qquad i=1,...,m+n.
$$
For integrable dominant weights, the linear combination has non-negative integer coefficients, up to two real numbers. The first  corresponds to the fermionic root, i.e. to $\lambda_m$.
The second  is associated to the eigenvalue of the $\mathfrak{gl}(1)$ part that distinguishes $\mathfrak{gl}(m|n)$ from its simple part $\mathfrak{sl}(m|n)$. It can be related to the weight $\boldsymbol{\lambda}^{(m+n)}$.

The representations associated to fundamental weights are called fundamental representations.
There are $m+n-1$ of them, and the first one, $\boldsymbol{\lambda}^{(1)}$ corresponds to what is usually called the fundamental representation.
It is $(m+n)$-dimensional, and in that case $\pi_{\boldsymbol{\lambda}^{(1)}}(\se_{ij})=E_{ij}$.
 Its contragredient representation (which is also $(m+n)$-dimensional) corresponds to $\boldsymbol{\lambda}^{(m+n-1)}$.

\subsection{Evaluation map\label{sec:eval}}
The \textit{evaluation morphism} $ev(\xi)$, for $\xi\in \mathbf{C}$, is an algebra morphism from $Y\bigl(\mathfrak{gl}(m|n)\bigr)$ to $U(\mathfrak{gl}(m|n))$, the enveloping algebra of $\mathfrak{gl}(m|n)$. It is defined by
\begin{equation}
ev(\xi):\quad T(u)\ \to\ \mathbf{I}+\frac{c}{u-\xi}\mathbf{E} \mb{with} \mathbf{E}=\sum_{i,j=1}^{m+n} (-1)^{[i]} E_{ij}\otimes \se_{ji},
\end{equation}
 with $\mathbf{I}=\mathbf{1}\otimes\textsf{1}$, where  we introduced \texttt{1} the unit of $U(\mathfrak{gl}(m|n))$ and we used the same notation as in section \ref{S-BN}. In component, the evaluation map reads
$$
ev(\xi)\big(T_{ij}(u)\big)= \delta_{ij}\,\textsf{1}+\frac{c_{[i]}}{u-\xi}\,\se_{ji}.
$$
Indeed, since the Lie superalgebra relations \eqref{com-glmn} are equivalent to
$$
[\mathbf{E}_1\,,\,\mathbf{E}_2] = P(\mathbf{E}_1-\mathbf{E}_2)\,,
$$
 it is easy to show that $\mathbf{I}+\frac{c}{u-\xi}\mathbf{E}$ obeys the Yangian $RTT$-relations \eqref{RTT}.
Remark that the generators of $\mathfrak{gl}(m|n)$ are related to the zero modes described in \cite{HutLPRS17b}: $\se_{ij}=(-1)^{[j]}\,T_{ji}[0]$.

Then, using the evaluation morphism one can construct, from any $\mathfrak{gl}(m|n)$ representation $\pi_{\boldsymbol{\lambda}}$,  a representation for the Yangian $Y\bigl(\mathfrak{gl}(m|n)\bigr)$.
The \textit{evaluation representation} $ev_{\boldsymbol{\lambda}}(\xi)=\pi_{\boldsymbol{\lambda}}\, o\, ev(\xi)$ is defined as:
$$
ev_{\boldsymbol{\lambda}}(\xi)\Big(T_{ij}(u)\Big)=
\delta_{ij}\textsf{1}_{\boldsymbol{\lambda}}+\frac{c_{[i]}}{u-\xi}\,e_{ji}\,,
$$
where $e_{ij}=\pi_{\boldsymbol{\lambda}}(\se_{ij})$ is the matrix representation of $\se_{ij}$ in the vector space
$\cal V_{\boldsymbol{\lambda}}$ and $\textsf{1}_{\boldsymbol{\lambda}}$ is the identity matrix in this space.
The weights of the Yangian representation $ev_{\boldsymbol{\lambda}}(\xi)$ read
$$
T_{ii}(u)|0\rangle = \lambda_{i}(u)|0\rangle \quad\mbox{with}\quad  \lambda_{i}(u)=1+\frac{c_{[i]}}{u-\xi}\,\lambda_i,
$$
 and we have
$$
T_{ij}(u)|0\rangle =\frac{c_{[i]}}{u-\xi}\,e_{ji}|0\rangle=0,\quad j<i
$$
according to the relations \eqref{hwglmn}.
Then it is clear that the highest weight vector of $\mathfrak{gl}(m|n)$ becomes the pseudovacuum vector \eqref{Tij}.

Let us emphasize the difference between $\lambda_i$, that are the weights for the Lie superalgebra $\mathfrak{gl}(m|n)$, and $\lambda_i(u)$, that are
the weights for the Yangian
$Y\bigl(\mathfrak{gl}(m|n)\bigr)$.

\subsection{Representations associated to $f_{[i]}(u,v)$}
For any $j=1,2,...,m+n$ and any complex $\xi$, we introduce the evaluation representation $Ev_{j}(\xi)$ associated to
the weight $\boldsymbol{\lambda}^{(j)}$. It  correspond to the Yangian weights
$$
\lambda_{\mu}(u)=\begin{cases} f_{[\mu]}(u,\xi) \quad &\mbox{ if } \mu\leq j, \\[1ex]
1\quad &\mbox{ if } \mu> j.\end{cases}
$$
We consider the following representation:
$\displaystyle\otimes_{j=1}^N\otimes_{k=1}^{L^{(j)}} Ev_j(\xi_k^{(j)})$. Since we have a tensor product of highest weight representations, the  weights for this tensor product are given by
the product of the individual weights for each representations, that is
$$
\lambda_{\mu}(u)=\prod_{j=\mu}^N\prod_{k=1}^{L^{(j)}} f_{[\mu]}(u,\xi_k^{(j)})\,,\quad \mu=1,2,...,m+n.
$$
This leads to \eqref{alpha-rep}.

\section{Recursion for the highest coefficient\label{S-RHC}}

One can build the HC $Z^{m|n}$ starting from the known results at $m+n=2$ via  recursions derived in \cite{HutLPRS17b}.
For $m=2$, $n=0$ we deal with the HC of $\mathfrak{gl}(2)$ based models, that is equal to the partition function of the six-vertex
model with domain wall boundary condition \cite{Kor82,Ize87}. The case $m=0$, $n=2$ becomes equivalent to the previous one after the
replacement the constant $c\to -c$ in the $R$-matrix \eqref{R-mat}. Finally, for $m=n=1$ the HC has the form \cite{HutLPRS16c}
\be{Z11}
Z^{1|1}(\bs|\bt)=g(\bs,\bt).
\ee

In recursive construction of the HC, two cases should be distinguished: (1) $n>0$ and $m>0$; (2) $n=0$ or $m=0$. We first
consider the case $n>0$ and $m>0$. Then, the recursive procedure is based on the following reductions  \cite{HutLPRS17b}:
\be{Rec-empty}
\begin{aligned}
&Z^{m|n}(\emptyset,\bs^2,\dots,\bs^{N}|\emptyset,\bt^2,\dots,\bt^{N})=Z^{m-1|n}(\bs^2,\dots,\bs^{N}|\bt^2,\dots,\bt^{N}),\\
&Z^{m|n}(\bs^1,\dots,\bs^{N-1},\emptyset|\bt^1,\dots,\bt^{N-1},\emptyset)=Z^{m|n-1}(\bs^1,\dots,\bs^{N-1}|\bt^1,\dots,\bt^{N-1}),
\end{aligned}
\ee
and we recall that $N=m+n-1$.
Thus, in particular, knowing $Z^{m-1|n}$ for some $m$ and $n$ we automatically know $Z^{m|n}$ with $\#\bs^1=\#\bt^1=0$. Then,
to obtain $Z^{m|n}$ with $\#\bs^1=\#\bt^1>0$ we can use a recursion \cite{HutLPRS17b}
\begin{multline}\label{Rec-HC1}
Z^{m|n}(\bs|\bt)
=\sum_{\rho=2}^{N+1}\sum_{\substack{\text{\rm part}(\bs^2,\dots,\bs^{\rho-1})\\\text{\rm part}(\bt^1,\dots,\bt^{\rho-1})}}
Z^{m|n}(\{\bs^\sigma_{\st}\}_1^{\rho-1},\{\bs^{\sigma}\}_\rho^N|\{\bt^\sigma_{\st}\}_1^{\rho-1},\{\bt^{\sigma}\}_\rho^N)
\left(\frac{g(\bs^1_{\st},\bs^1_{\so})}{f(\bs^1_{\st},\bs^1_{\so})}\right)^{\delta_{m,1}}\\
\times \frac{g_{[2]}(\bt^{1}_{\so},\bs^{1}_{\so})
\gamma_{1}(\bt^{1}_{\so},\bt^{1}_{\st})f(\bt^{1}_{\st},\bs^{1}_{\so})}
{f_{[\rho]}(\bar s^{\rho},\bar s^{\rho-1}_{\so})}
\prod_{\nu=2}^{\rho-1}\frac{
g_{[\nu+1]}(\bt^{\nu}_{\so},\bt^{\nu-1}_{\so})g_{[\nu]}(\bs^\nu_{\so},\bs^{\nu-1}_{\so})
\gamma_{\nu}(\bt^{\nu}_{\so},\bt^{\nu}_{\st})\gamma_{\nu}(\bs^\nu_{\st},\bs^{\nu}_{\so}) }
{f_{[\nu]}(\bar s^{\nu},\bar s^{\nu-1}_{\so})
f_{[\nu]}(\bar t^{\nu}_{\so},\bar t^{\nu-1})}\;.
\end{multline}
Here
\begin{multline}\label{Zdet1}
Z^{m|n}(\{\bs^\sigma_{\st}\}_1^{\rho-1},\{\bs^{\sigma}\}_\rho^N|\{\bt^\sigma_{\st}\}_1^{\rho-1},\{\bt^{\sigma}\}_\rho^N)\\
=Z^{m|n}(\bs^1_{\st},\dots,\bs^{\rho-1}_{\st},\bs^{\rho},\dots,\bs^{N}|
\bt^1_{\st},\dots,\bt^{\rho-1}_{\st},\bt^{\rho},\dots,\bt^{N}).
\end{multline}
For    every  fixed $\rho\in\{2,\dots,N+1\}$  in \eqref{Rec-HC1}
the sums are taken over partitions $\bt^\sigma\Rightarrow\{\bt^\sigma_{\so}, \bt^\sigma_{\st} \}$
with $\sigma=1,\dots,\rho-1$
and $\bs^\sigma\Rightarrow\{\bs^\sigma_{\so}, \bs^\sigma_{\st} \}$ with $\sigma=2,\dots,\rho-1$, such that $\#\bt^\sigma_{\so}=\#\bs^\sigma_{\so}=1$. The subset $\bs^1_{\so}$
is a fixed Bethe parameter from the set $\bs^1$. There is no sum over partitions of the set $\bs^1$ in \eqref{Rec-HC1}.

Similarly, knowing $Z^{m|n-1}$ for some $m$ and $n$ we automatically know
$Z^{m|n}$ with $\#\bs^{N}=\#\bt^{N}=0$. Then, to obtain $Z^{m|n}$ with $\#\bs^{N}=\#\bt^{N}>0$ we can use the second recursion
\begin{multline}\label{Rec-2HC1}
Z^{m|n}(\bs|\bt)=\sum_{\rho=1}^{N}
\sum_{\substack{\text{\rm part}(\bs^\rho,\dots,\bs^{N})\\\text{\rm part}(\bt^\rho,\dots,\bt^{N-1})}}
Z^{m|n}(\bigl\{\bs^\sigma\bigr\}_1^{\rho-1},\bigl\{\bs^{\sigma}_{\st}\bigr\}_\rho^N|
\bigl\{\bt^\sigma\bigr\}_1^{\rho-1};\bigl\{\bt^\sigma_{\st}\bigr\}_\rho^{N})
\left(\frac{g(\bt^N_{\st},\bt^N_{\so})}{f(\bt^N_{\st},\bt^N_{\so})}\right)^{\delta_{m,N}}\\
\times\frac{g(\bs^{N}_{\so},\bt^{N}_{\so})\gamma_{N}(\bs^{N}_{\st},\bs^{N}_{\so})
f(\bs^{N}_{\st},\bt^{N}_{\so})}{f_{[\rho]}(\bt^{\rho}_{\so},\bt^{\rho-1})}
\prod_{\nu=\rho}^{N-1}\frac{g_{[\nu]}(\bs^{\nu+1}_{\so},\bs^{\nu}_{\so})g_{[\nu]}(\bt^{\nu+1}_{\so},\bt^{\nu}_{\so})
\gamma_{\nu}(\bs^{\nu}_{\st},\bs^{\nu}_{\so})\gamma_{\nu}(\bt^{\nu}_{\so},\bt^{\nu}_{\st})}
{f_{[\nu+1]}(\bs^{\nu+1},\bs^{\nu}_{\so})f_{[\nu+1]}(\bar t^{\nu+1}_{\so},\bar t^{\nu})}\;.
\end{multline}
Here
\begin{multline}\label{Zdet2}
Z^{m|n}(\bigl\{\bs^\sigma\bigr\}_1^{\rho-1},\bigl\{\bs^{\sigma}_{\st}\bigr\}_\rho^N|
\bigl\{\bt^\sigma\bigr\}_1^{\rho-1};\bigl\{\bt^\sigma_{\st}\bigr\}_\rho^{N})\\
=Z^{m|n}(\bs^1,\dots,\bs^{\rho-1},\bs^{\rho}_{\st},\dots,\bs^{N}_{\st}|
\bt^1,\dots,\bt^{\rho-1},\bt^{\rho}_{\st},\dots,\bt^{N}_{\st}).
\end{multline}
For    every  fixed $\rho\in\{1,\dots,N\}$ in \eqref{Rec-2HC1} the sums are taken over partitions
$\bt^\sigma\Rightarrow\{\bt^\sigma_{\so}, \bt^\sigma_{\st} \}$ with $\sigma=\rho,\dots,N-1$
and $\bs^\sigma\Rightarrow\{\bs^\sigma_{\so}, \bs^\sigma_{\st} \}$ with $\sigma=\rho,\dots,N$,
such that $\#\bt^\sigma_{\so}=\#\bs^\sigma_{\so}=1$. The subset $\bt^N_{\so}$
is a fixed Bethe parameter from the set $\bt^N$. There is no sum over partitions of the set $\bt^N$ in \eqref{Rec-2HC1}.

Now, let us describe the situation in the case $n=0$.
The formulas \eqref{Rec-HC1}, \eqref{Rec-2HC1} remain valid in this case, however, they are slightly simplified. First of all,
$\delta_{m,1}=\delta_{m,N}=0$ in this case. This leads to the disappearance of the factors in the first lines of \eqref{Rec-HC1}, \eqref{Rec-2HC1}.
Second, all the $\gamma$-functions should be replaced by the $f$-functions.
Finally, all the subscripts of the $g$-functions and $f$-functions disappear: $g_{[\nu]}(x,y)\to g(x,y)$, $f_{[\nu]}(x,y)\to f(x,y)$.

However, the main peculiarity of this case is that the reductions \eqref{Rec-empty} take the form
\be{Rec-empty0}
\begin{aligned}
&Z^{m|0}(\emptyset,\bs^2,\dots,\bs^{m-1}|\emptyset,\bt^2,\dots,\bt^{m-1})=Z^{m-1|0}(\bs^2,\dots,\bs^{m-1}|\bt^2,\dots,\bt^{m-1}),\\
&Z^{m|0}(\bs^1,\dots,\bs^{m-2},\emptyset|\bt^1,\dots,\bt^{m-2},\emptyset)=Z^{m-1|0}(\bs^1,\dots,\bs^{m-2}|\bt^1,\dots,\bt^{m-2}).
\end{aligned}
\ee
Thus, if either $\bs^1=\bt^1=\emptyset$ or $\bs^{m-1}=\bt^{m-1}=\emptyset$, then in both cases $Z^{m|0}$ reduces
to $Z^{m-1|0}$.

Finally, the case of $\mathfrak{gl}(0|n)$ algebras reduces to the case considered above  after the
replacement the constant $c\to -c$ in the $R$-matrix \eqref{R-mat}. Therefore, we do not consider this case below.

\section{Residues in the poles of the highest coefficient\label{SS-RP}}

We give a detailed proof of proposition~\ref{Res-HC} for the case $m>0$ and $n>0$. The case $m=0$ or $n=0$ can be considered exactly in the same manner.

The proof is based on the reductions \eqref{Rec-empty}, recursions  \eqref{Rec-HC1}, \eqref{Rec-2HC1}, and explicit
representation \eqref{Z11} for $Z^{1|1}(\bs|\bt)$. First, one can easily see that due to \eqref{Z11}
\be{Z11-res}
Z^{1|1}(\bs|\bt)\Bigr|_{s_j\to t_j}=g(s_j,t_j)g(\bs_j,s_j)g(t_j,\bt_j)Z^{1|1}(\bs_j|\bt_j)+reg.
\ee
This expression obviously coincides with \eqref{res-otv} for $m=n=1$. Equation \eqref{Z11-res} serves as the basis of induction\footnote{%
For completeness of the proof one should also check \eqref{res-otv} for $m=2$ and $n=0$. This was done in \cite{Kor82}.}.

Assume that \eqref{res-otv} is valid  for all $m'$ and $n'$, such that
$m'+n'$ is fixed.  Then due to \eqref{Rec-empty} the residue formula \eqref{res-otv} holds for $Z^{m|n}$ with $m=m'+1$, $n'=n$ at $r_1=0$ (that is, $\bs^1=\bt^1=\emptyset$) and
for $Z^{m|n}$ with $m=m'$, $n=n'+1$ at $r_{N}=0$ (that is, $\bs^{N}=\bt^{N}=\emptyset$). Then using recursions \eqref{Rec-HC1} and \eqref{Rec-2HC1} we should prove that \eqref{res-otv} remains true for $r_1>0$ and $r_{N}>0$. It so happens that recursion \eqref{Rec-HC1} allows one to prove \eqref{res-otv} for
$\bs^\mu$ and $\bt^\mu$ with $\mu=2,\dots,N$. At the same time recursion \eqref{Rec-2HC1} provides the proof for  $\bs^\mu$ and $\bt^\mu$ with $\mu=1,\dots,N-1$. Combining both recursions we prove the residue formula \eqref{res-otv} for  all $\bs^\mu$ and $\bt^\mu$.

Let us show how this method works. Consider, for example, the recursion \eqref{Rec-HC1}. It is convenient to write
it in the following form:
\be{Rec1-short}
Z^{m|n}(\bs|\bt)=\sum_{\rho=2}^{N+1}\mathcal{Z}_\rho^{m|n}(\bs|\bt),
\ee
where
\begin{multline}\label{Rec-HC1rho}
\mathcal{Z}_\rho^{m|n}(\bs|\bt)
=\sum_{\substack{\text{\rm part}(\bs^2,\dots,\bs^{\rho-1})\\\text{\rm part}(\bt^1,\dots,\bt^{\rho-1})}}
Z^{m|n}(\{\bs^\sigma_{\st}\}_1^{\rho-1},\{\bs^{\sigma}\}_\rho^N|\{\bt^\sigma_{\st}\}_1^{\rho-1},\{\bt^{\sigma}\}_\rho^N)
\left(\frac{g(\bs^1_{\st},\bs^1_{\so})}{f(\bs^1_{\st},\bs^1_{\so})}\right)^{\delta_{m,1}}\\
\times \frac{g_{[2]}(\bt^{1}_{\so},\bs^{1}_{\so})
\gamma_{1}(\bt^{1}_{\so},\bt^{1}_{\st})f(\bt^{1}_{\st},\bs^{1}_{\so})}
{f_{[\rho]}(\bar s^{\rho},\bar s^{\rho-1}_{\so})}
\prod_{\nu=2}^{\rho-1}\frac{
g_{[\nu+1]}(\bt^{\nu}_{\so},\bt^{\nu-1}_{\so})g_{[\nu]}(\bs^\nu_{\so},\bs^{\nu-1}_{\so})
\gamma_{\nu}(\bt^{\nu}_{\so},\bt^{\nu}_{\st})\gamma_{\nu}(\bs^\nu_{\st},\bs^{\nu}_{\so}) }
{f_{[\nu]}(\bar s^{\nu},\bar s^{\nu-1}_{\so})
f_{[\nu]}(\bar t^{\nu}_{\so},\bar t^{\nu-1})}\;.
\end{multline}
We first consider the case $r_1=\#\bs^1=\#\bt^1=1$. Then $\#\bs^1_{\st}=\#\bt^1_{\st}=0$, hence, we actually have $Z^{m-1|n}$ in the rhs
of \eqref{Rec-HC1rho}. According to the induction assumption the residue formula \eqref{res-otv} is valid for these HC.


Let $s_j^\mu=t_j^\mu$ for $\mu>1$ in the lhs of \eqref{Rec1-short}. In the rhs of this equation one should consider separately the terms with
different $\rho$. Namely, one should distinguish between four cases:  $\rho<\mu$; $\rho > \mu+1$; $\rho=\mu+1$; $\rho=\mu$.

Let $\rho<\mu$. The pole at $s_j^\mu=t_j^\mu$ in the rhs of \eqref{Rec-HC1rho} occurs in the HC only.
Then due to the induction assumption the residue of the HC in the rhs of \eqref{Rec-HC1rho} gives the factor
\be{coef1}
\mathcal{A}_\mu=\frac{g_{[\mu+1]}(t_j^\mu,s_j^\mu)\gamma_{\mu}(\bt_j^\mu,t_j^\mu)\gamma_{\mu}(s_j^\mu,\bs_j^\mu) } {f_{[\mu+1]}(\bt^{\mu+1},t_j^{\mu})f_{[\mu]}(s_j^{\mu},\bs^{\mu-1})}.
\ee
This coefficient does not depend on the partitions. The remaining sum over partitions
obviously reduces to $\mathcal{Z}_\rho^{m|n}(\bs\setminus\{s^\mu_j\}|\bt\setminus\{t^\mu_j\})$. Thus, for $\rho<\mu$ we arrive at
\be{1CASE}
\mathcal{Z}_\rho^{m|n}(\bs|\bt)\Bigr|_{s^\mu_j=t^\mu_j}
=\mathcal{A}_\mu\mathcal{Z}_\rho^{m|n}(\bs\setminus\{s^\mu_j\}|\bt\setminus\{t^\mu_j\})+reg.
\ee

Consider now the terms with $\rho> \mu+1$. The pole in the rhs of \eqref{Rec-HC1rho} occurs in the HC provided $s^\mu_j\in\bs^\mu_{\st}$ and
$t^\mu_j\in\bt^\mu_{\st}$. Let $\bs^\mu_{\st}=\{s_j^\mu,\bs^\mu_{\st'}\}$ and $\bt^\mu_{\st}=\{t_j^\mu,\bt^\mu_{\st'}\}$. Then
the residue of the highest coefficient gives the factor
\be{coef2}
\frac{g_{[\mu+1]}(t_j^\mu,s_j^\mu)\gamma_{\mu}(\bt_{\st'}^\mu,t_j^\mu)\gamma_{\mu}(s_j^\mu,\bs_{\st'}^\mu)}
{f_{[\mu+1]}(\bt^{\mu+1}_{\st},t_j^{\mu})f_{[\mu]}(s_j^{\mu},\bs^{\mu-1}_{\st})}.
\ee
The second line of \eqref{Rec-HC1rho} gives additional factors depending on $s^\mu_j$ and $t^\mu_j$:
\be{coeff-2a}
\frac{\gamma_{\mu}(\bt^{\mu}_{\so},t^{\mu}_{j})\gamma_{\mu}(s^\mu_{j},\bs^{\mu}_{\so})}
{f_{[\mu+1]}(\bt^{\mu+1}_{\so},t_j^{\mu})f_{[\mu]}(s_j^{\mu},\bs^{\mu-1}_{\so})} . 
\ee
Together with \eqref{coef2} they give $\mathcal{A}_\mu$ \eqref{coef1}. The rest of \eqref{Rec-HC1rho} does not depend on
$s^\mu_j$ and $t^\mu_j$, hence, we again obtain \eqref{1CASE}, but now for $\rho> \mu+1$.
%
%

The third case is $\rho=\mu+1$. Again, the pole occurs in the HC, and we set $\bs^\mu_{\st}=\{s_j^\mu,\bs^\mu_{\st'}\}$, $\bt^\mu_{\st}=\{t_j^\mu,\bt^\mu_{\st'}\}$. Now the factor coming from the HC is
\be{coef3}
\frac{g_{[\mu+1]}(t_j^{\mu},s_j^{\mu})\gamma_{\mu}(\bt_{\st'}^{\mu},t_j^{\mu})\gamma_{\mu}(s_j^{\mu},\bs_{\st'}^{\mu}) }
{f_{[\mu+1]}(\bt^{\mu+1},t_j^{\mu})f_{[\mu]}(s_j^{\mu},\bs^{\mu-1}_{\st})}.
\ee
We also have from the second line of \eqref{Rec-HC1rho}
\be{coeff-3a}
\frac{\gamma_{\mu}(\bt^{\mu}_{\so},t^{\mu}_{j})\gamma_{\mu}(s^{\mu}_{j},\bs^{\mu}_{\so}) }
{f_{[\mu]}(s_j^{\mu},\bs^{\mu-1}_{\so})},
\ee
and altogether we again obtain \eqref{coef1}. Thus, equation \eqref{1CASE} holds for $\rho=\mu+1$.

Finally, let $\rho=\mu$. Then we have form the HC
\be{coef4}
\frac{g_{[\mu+1]}(t_j^{\mu},s_j^{\mu})\gamma_{\mu}(\bt_{j}^{\mu},t_j^{\mu})\gamma_{\mu}(s_j^{\mu},\bs_{j}^{\mu})}
{f_{[\mu+1]}(\bt^{\mu+1},t_j^{\mu})f_{[\mu]}(s_j^{\mu},\bs^{\mu-1}_{\st})} .
\ee
The additional factor $f_{[\mu]}(s_j^{\mu},\bs^{\mu-1}_{\so})$ comes from the second line of \eqref{Rec-HC1rho}, and we again
obtain the $\mathcal{A}_\mu$ coefficient \eqref{coef1}.
 The remaining sum over partitions still gives
$\mathcal{Z}_\rho^{m|n}(\bs\setminus\{s^\mu_j\}|\bt\setminus\{t^\mu_j\})$.

Thus, equation \eqref{1CASE} is proved for all $\rho$. Due to \eqref{Rec1-short} this immediately yields the residue formula \eqref{res-otv} for
$\mathcal{Z}^{m|n}$.

As soon as \eqref{res-otv} is proved for $r_1=1$ we can use it as a new basis of induction. We assume that  \eqref{res-otv} is valid for some $r_1>0$
and then prove that it remains true for $r_1+1$. All considerations are exactly the same as in the case $r_1=1$, therefore we omit them.

In this way we prove the residue formula for all $\bs^\mu$ and $\bt^\mu$ except $\bs^1$ and $\bt^1$. To prove \eqref{res-otv} for the
residue at $s^1_j=t^1_j$ we should use the second recursion \eqref{Rec-2HC1} and perform similar calculations.


\end{document}